 \documentclass[preprint2]{aastex63}
\received{September 19, 2020}
\revised{January 1, 2021}
\accepted{February 2, 2021}
\submitjournal{ApJ}
\shorttitle{WCCC and irradiation}
\shortauthors{Kalv\=ans}

\begin{document}

\title{The connection between warm carbon chain chemistry and interstellar irradiation of star-forming cores}

\correspondingauthor{Juris Kalv\=ans}
\email{juris.kalvans@venta.lv}

\author[0000-0002-2962-7064]{Juris Kalv\=ans}
\affiliation{Ventspils University of Applied Sciences \\
Engineering Research Institute ``Ventspils International Radio Astronomy Center'' \\
In$\rm \check{z}$enieru 101, Ventspils, LV-3601, Latvia}

\begin{abstract}
Some observations of warm carbon chain chemistry (WCCC) cores indicate that they are often located near the edges of molecular clouds. This finding may suggest that WCCC is promoted in star-forming cores exposed to radiation from the interstellar medium. We aim to investigate the chemistry of carbon chains in such a core. A chemical simulation of a gas parcel in a low-mass star-forming core with a full level of irradiation by interstellar photons and cosmic rays was compared to a simulation of a core receiving only one-tenth of such irradiation. In the full irradiation model, the abundances of carbon chains were found to be higher by a factor of few to few hundred, compared to the model with low irradiation. Higher carbon-chain abundances in the prestellar stage and, presumably, in the extended circumstellar envelope, arise because of irradiation of gas and dust by interstellar photons and cosmic rays. A full standard rate of cosmic-ray induced ionization is essential for a high carbon-chain abundance peak to occur in the circumstellar envelope, which is heated by the protostar (the ``true'' WCCC phenomenon). The full irradiation model has lower abundances of complex organic molecules than the low-irradiation model. We conclude that WCCC can be caused by exposure of a star-forming core to interstellar radiation, or even just to cosmic rays. The Appendix describes an updated accurate approach for calculating the rate of cosmic-ray induced desorption.
\end{abstract}

%\keywords{Astrochemistry --- Interstellar abundances --- Interstellar medium --- Dense interstellar clouds --- Collapsing clouds --- Star formation --- Interstellar molecules --- Interstellar dust processes --- Cosmic rays}

\section{Introduction}
\label{intrd}

Warm carbon-chain chemistry (WCCC) in low-mass young stellar objects (YSOs) was first discovered by \citet{Sakai08}. The envelope in the L1527 molecular cloud hosting an embedded Class 0 or 0/I protostar IRAS 04368+2557 \citep{Ohashi97,Jorgensen02,Sakai13} was found to be rich in carbonaceous species, such as C$_2$H, c-C$_3$H$_2$, C$_4$H, and others. The beam-averaged temperature was found to be 14\,K, while the density is $\approx$10$^6$\,cm$^{-3}$. Thus, the observations of protostellar carbon chains were associated with dense ``lukewarm'' (gas temperature $\lesssim$40\,K) circumstellar envelopes. WCCC is distinct from the hot corino chemistry, characterized by abundant and warm ($>$100\,K) oxygen-containing complex organic molecules \citep[COMs;][]{Cazaux03,Bottinelli04}, such as methanol CH$_3$OH.

The discovery of additional sources, such as IRAS 15398--3359 in Lupus and IRAS 18148--0440 in L483, established WCCC as a separate type of YSOs \citep{Hirota09,Sakai09lupus,Sakai16,Cordiner11,Graninger16,Higuchi18}. Other observations revealed a diverse variety of carbon chains \citep{Sakai07,Sakai08-,Sakai09deut,Hirota10,Tokudome13,Agundez15,Araki16,Yoshida19}. The sources were found to consist of a WCCC core, rich in smaller molecules, such as c-C$_3$H$_2$, and an extended circumstellar envelope, more rich in longer chains, such as C$_6$H$_2$ and C$_7$H, probably remnants from the prestellar cloud core. Other molecules, such as  C$_2$H and C$_4$H, are present in both components. The WCCC molecules were found to peak slightly offset relative to the protostar, i.e., in the lukewarm regions of the circumstellar envelope \citep{Sakai09lupus,Sakai10,Araki17}. ALMA observations show that the very center of the core may contain COMs \citep{Jorgensen13,Imai16,Oya17}.

WCCC or similar sources have been found to be associated with very low luminosity objects \citep[YSOs in the earliest star-forming stages;][]{Takakuwa11,Cordiner12cha} as well as evolved protostars \citep{Oya17}. Thus, the carbon-chain rich chemistry can be attributed to starless clouds, such as TMC-1 \citep[e.g.,][]{Takano90,Sakai07}, first hydrostatic cores, and Class 0 and I protostars. Later evolutionary stages are likely to show hot-corino type chemistry \citep[i.e., rich in hot COMs,][]{Graninger16}. Even massive star-forming regions, rich in carbon chains, have been found \citep{Mookerjea10,Mookerjea12,Saul15,Taniguchi18mas}. A special case of a WCCC object might be the L1489 molecular cloud, which is heated externally by the nearby protostar L1489 IRS \citep{Wu19aa}. Warm carbon chains have also been observed in shocked regions of the interstellar medium \citep[ISM;][]{Lefloch18,Wu19mn}.

Observational surveys of protostars show that hot corinos and WCCC cores probably are two extremes and that many YSOs are intermediate in their content of COMs and carbon chains \citep{Graninger16,Lindberg16,Higuchi18,Law18}. A number of high-resolution interferometric observations have studied the inner workings of the central engines of WCCC cores \citep[e.g.,][]{Sakai14,Sakai14nat,Sakai17,Aso15,Oya15}, revealing that infalling matter undergoes a mild shock at the transition zone between the envelope and the protostellar disk. There is a lack of medium spatial resolution interferometric observations that are sensitive to the spatial scale of a circumstellar envelope \citep{Oya20}.

The working hypothesis for the origin of low-mass WCCC protostellar cores is gas-phase processing of methane CH$_4$, evaporated from ices in the lukewarm regions of the circumstellar envelope  \citep{Sakai08,Sakai09lupus}. This picture was immediately supported by several astrochemical modeling studies \citep{Aikawa08,Harada08,Hassel08} and is supported also by detections of methane and methane ice in YSOs rich in carbon-chains \citep{Oberg08,Sakai12}. Additionally, models also indicate that part of the carbon chain inventory in WCCC cores has been inherited from a previous starless phase \citep{Aikawa01,Hassel08,Cordiner12-}. The presence of such remnant molecules is co-opted by the observations of rich carbon-chain chemistry in early YSOs \citep{Takakuwa11}. Even more so, \citet{Cordiner12-} argue that WCCC is promoted by the depletion of oxygen onto the surfaces of grains and the mediation of CH$_4$ ice is unnecessary. Their interpretation of observations also support the idea that carbon chains have been left over from the prestellar phase \citep{Charnley10,Cordiner12cha}.

In line with observational evidence, models show that WCCC and hot corino chemistry (i.e., abundant COMs in temperatures above 100\,K) can coexist in a single source \citep{Hassel11,Garrod17,Aikawa20}. The WCCC is initiated by the warm-up of the protostellar envelope, inducing evaporation of methane, which is converted to carbon chains and similar species via gas phase chemistry. As the protostar heats the envelope and the temperature rises, these species are synthesized, frozen-out and evaporated again \citep{Taniguchi19}. Not all of the CH$_4$ ice might be available for immediate sublimation and participation in WCCC because the watery ice matrix may partially lock CH$_4$ inside the mantle \citep{Wang19,Aikawa20}.

A less clear question is the source of CH$_4$ ice, necessary to produce the WCCC cores. \citet{Sakai08,Sakai09lupus,Sakai09deut} suggested that a faster collapse of the prestellar core would not allow CO molecules to become the dominant form of carbon, instead carbon would accumulate onto grains as CH$_4$ ice. This idea was apparently disproved by the model of \citet{Aikawa20}. Astrochemical models do not indicate such a possibility of rapid CH$_4$ synthesis, as CO always seems to dominate the C budget in a gas with hydrogen numerical density $n_H$ exceeding about $10^4$\,cm$^{-3}$, especially when CO self-shielding and mutual shielding by H$_2$ from the interstellar radiation field (ISRF) are considered \citep[e.g.,][]{Herbst89,Aikawa01,Rawlings02,Garrod06}. The process that generates methane ice must be uncommon, as only relatively few star-forming cores are affected by it.

Observations give clues to the origin of WCCC. Carbon-chain rich cores in various evolutionary stages reside in specific regions in the ISM, such as Heiles and Lupus~I clouds. Additionally, the surveys by \citet{Higuchi18} and \citet{Lefloch18} indicate that WCCC cores tend to be located outside dense concentrations and thus are likely to be exposed to higher radiation levels than non-WCCC and hot corino sources. The possible association of WCCC cores with higher radiation is also evidenced by WCCC observations near massive star-forming regions \citep{Taniguchi18poly,Sicilia19} and Herbig Ae/Be stars \citep{Lindberg12,Lindberg16,Watanabe12}. Exposure to an increased radiation flux causes high abundances of carbon chains also in starless and prestellar cores \citep[e.g.,][]{Li16,Spezzano16,Pan17}. The latter observation gives the possibility that high carbon chain abundances simply are the result of synthesis in irradiated cold gas. However, conditions in the WCCC cores have not been thoroughly measured and their connection to elevated radiation is not clear, meaning that irradiation is only one possible cause or prerequisite for WCCC.

The aim of this study is to qualitatively investigate the chemistry of carbon chains in star-forming cores irradiated from outside by comparing the chemistry of carbon chains in a shielded cloud core and in a core exposed to radiation. In this way, we aim to understand, if outside irradiation can facilitate the formation of a WCCC core.

Isolated cores outside larger dense structures are more exposed to the ISRF and cosmic rays (CRs). This results in an overall greater flux of ionizing radiation, even deep within the core. Moreover, CR-induced desorption (CRD), caused by whole-grain heating, has recently been found to be potentially very efficient \citep{K18aps}. Thus, CRD may significantly delay the freeze out of both, CO and CH$_4$ \citep{KK19}, affecting the possibility of WCCC. The observations by \citet{Bouvier20} show that intense irradiation by nearby newborn massive stars can turn the star-forming cloud into a photodissociation region (PDR), changing the balance between COMs and carbon-chains. In such a case, neither hot corino, nor WCCC activity was observed; the observed CCH and CH$_3$OH emission originated in the PDR. This means that we need to consider only mild increase of radiation, not the orders of magnitude increase experienced by PDRs.

In order to fulfill the above aim, we perform two simulations of chemistry in star-forming cores. One of these was assumed to have a modest irradiation, relevant to a shielded dense core, while the other was assumed to be exposed to a higher level of interstellar radiation. The astrochemical model is described below in Section~\ref{mthd}, while the results of the simulations are analyzed in Section~\ref{rslt}. The final conclusions were drawn in Section~\ref{cncl}.

\section{Methods}
\label{mthd}

We employ the astrochemical model \textsc{Alchemic-Venta}, initially developed by \citet{K15apj1}. The latest version of this model is described by \citet{KK19}. The model follows the chemistry in a gas parcel with conditions relevant to a collapsing dense cloud core that transforms into an initial, still spherical, protostellar envelope. In the envelope stage, the parcel undergoes heating by the central engine, while other conditions remain unchanged. A comprehensive description of the model follows.

\subsection{Physical model}
\label{mmacr}
%
% Figure 1
\begin{figure} [htb!]
%\centering
\vspace{3.5cm}
%\plotone{fig-chord.eps}
\hspace{1.0cm}
\includegraphics[width=15cm]{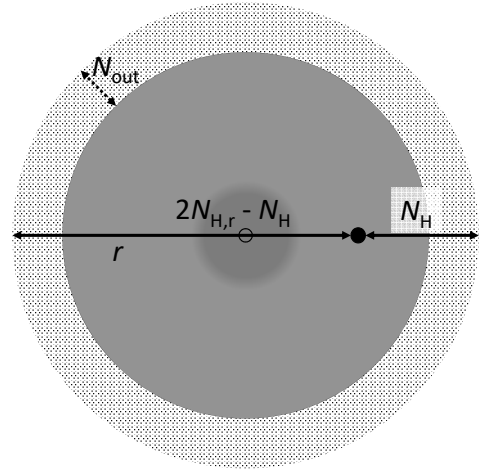}
\vspace{-8.0cm}
\caption{Schematic geometrical setting of the model (not to scale). The filled circle indicates the location of the modeled cloud parcel, while the solid black arrows indicate the segments of core diameter, along which the hydrogen column density $N_H$ was calculated. The empty circle indicates core center. \label{fig-rad}}
\end{figure}

The task of the physical model is to simulate the evolution of the specific gas parcel in a low-mass star-forming core that eventually hosts WCCC. The model is neither intended nor is able to simulate realistic evolution and geometry of the whole star-forming core. The macrophysical model considers a 1D (shell) structure of a spherical star-forming core in Lagrangian mass coordinates. We assumed a core mass of $M_{\rm core}$=1.0\,$M_{\odot}$. Chemistry was calculated only at one point of the model, in order to obtain clearly interpretable results. This point, or parcel of infalling gas, was chosen so that it ends up having a hydrogen number density of $\approx10^6$\,cm$^{-3}$, in line with observations of WCCC cores and other WCCC models. The parcel is located in a shell with a constant relative mass coordinate of $X_m$=0.55. The core collapse stage ends with the assumed formation of the protostar at $t$=1.6\,Myr. A 0.84\,Myr long protostellar stage follows, where the parcel density and column density towards the parcel were held constant, while temperature gradually rises, simulating the heating by the protostar.
%
% Table 1
\begin{table*}
\begin{center}
\caption{Physical parameters at the start and the assumed end of the 1.0\,M$_\odot$ prestellar core collapse stage. The indices 0, 0.55, and 1 indicate the relative mass coordinate $X_m$, except for $r_0$, which indicates the radius of the density plateau.}
\label{tab-macr}
\begin{tabular}{lccccccc}
\hline\hline\\
%\multicolumn{2}{l}{Gas density} & 
%\multicolumn{2}{c}{$a_{\rm disr}(\mu m)$}\\
$t$ (Myr) & $n_{H,0}$ (cm$^{-3}$) & $n_{H,0.55}$ (cm$^{-3}$) & $A_{V,0}$ (mag) & $A_{V,0.55}$ (mag) & $r_{0.55}$ (AU) & $r_0$ (AU) & $r_1$ (AU) \\
\hline
0 & $2.0\times10^3$ & $1.8\times10^3$ & 1.9 & 1.5 & $2.5\times10^4$ & $8.6\times10^4$ & $3.1\times10^4$ \\
1.602 & $3.7\times10^6$ & $1.0\times10^6$ & 68 & 9.0 & $2.6\times10^3$ & $2.3\times10^3$ & $3.8\times10^3$ \\
\hline
\end{tabular}
\end{center}
\end{table*}

The physical model for the prestellar core was built as follows. The hydrogen number density $n_{H,0}$ at the center of the core ($X_m$=0) evolves according to a free-fall collapse, increasing by three orders of magnitude (see Table~\ref{tab-macr}). The collapse is delayed by a factor of 0.7, as assumed by a number of authors, starting with \citet{Nejad90}. With an initial density of 2000\,cm$^{-3}$, such a delayed collapse occurs over an integration time of $t$$\approx$$1.6$\,Myr. The initial density was chosen relatively low to allow for chemical relaxation to reach a pseudo-equilibrium. It is important that the relaxation occurs before the accumulation of ices, whose composition affects chemistry in the protostellar stage.

With $n_{H,0}$ now known, the density $n_{H,r}$ at a core radius $r$ was then calculated with a Plummer-like approximation \citep{Plummer11,Whitworth01,Taquet14}
\begin{equation}
	\label{mmacr1}
	n_{H,r} = \frac{n_{H,0}} {(1 + \left(\frac{r}{r_0}\right)^2)]^{\eta/2}},
\end{equation}
where $r_0$ is the radius of the central density plateau, proportional to $n_{H,0}^{-1/2}$ \citep{Keto10} and $\eta$ was taken to be 3. Equation~(\ref{mmacr1}) was employed to calculate the gas density and mass of core shells with radius $r$, ranging from 0 to $r_1$. The latter value corresponds to $X_m$=1.0. In other words, the total radius of the core was maintained such that the core mass always is 1.0\,$M_{\odot}$. The core was assumed to be surrounded by gas with a column density of $N_{\rm out} = 2\times10^{21}$\,cm$^{-2}$, a feasible value for solar-mass cores \citep{Launhardt13}.

% Figure 2
\begin{figure} [htb!]
\vspace{-1.0cm}
%\plotone{fig-nhav.eps}
\hspace{-2.0cm}
\includegraphics[width=22cm]{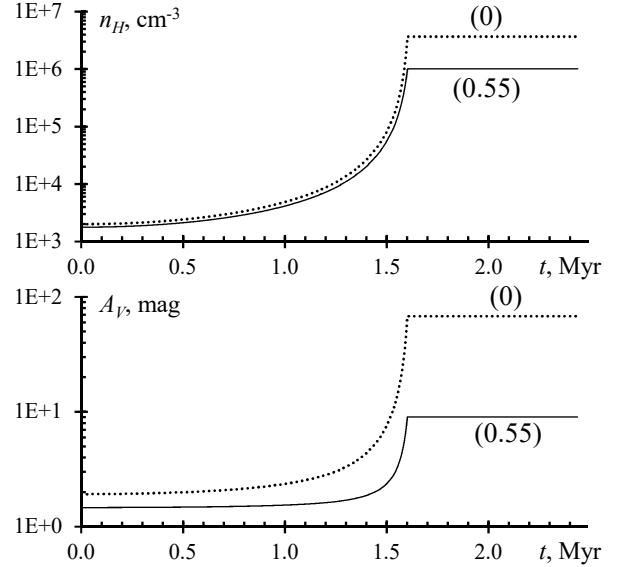}
\vspace{-22.0cm}
\caption{Evolution of key parameters for the simulation of chemistry in a star-forming cloud core. Top: $n_H$ at core center ($X_m$=0) and at the parcel located at $X_m$=0.55. Bottom: the $A_V$ values for the core center and the parcel. \label{fig-nhav}}
\end{figure}
%
% Figure 3
\begin{figure} [htb!]
\vspace{-1.0cm}
%\plotone{fig-phys.eps}
\hspace{-2.0cm}
\includegraphics[width=22cm]{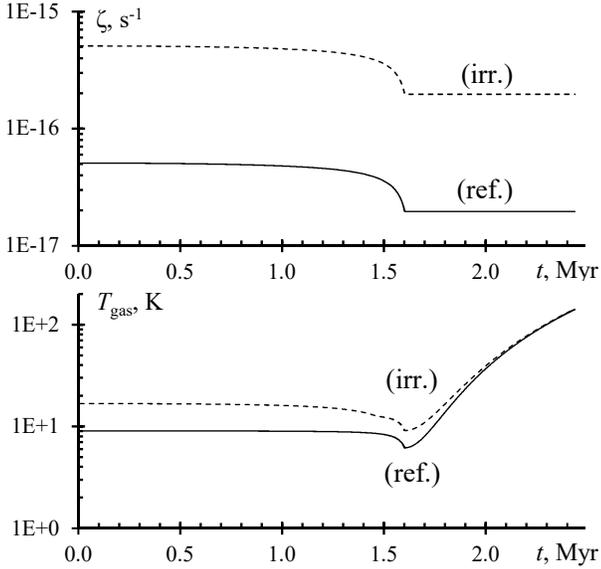}
\vspace{-22.0cm}
\caption{Evolution of the CR ionization rate $\zeta$ and gas temperature for the irradiation and reference models. Dust temperature follows a curve similar to that of $T_{\rm gas}$. \label{fig-phys}}
\end{figure}
The known 1D macroscopical structure of the cloud core was used to calculate column density $N_H$ and ISRF extinction $A_V$, assuming that $A_V=N_H/1.6\times10^{21}$. With the initial conditions listed in Table~\ref{tab-macr}, the initial visual extinction at $X_m$=0.55 is $A_{V,0.55}$=1.5\,mag, i.e., below the value when interstellar water ice has been observed accumulating on grain surfaces \citet[at a line-of-sight $A_V$ of 3.2\,mag]{Whittet01}. Figure~\ref{fig-rad} shows that the parcel was assumed to be irradiated from two opposite directions along the diameter of the core. The macrophysical model was designed so that gas density $n_{H,0.55}$ at $X_m$=0.55 is $10^6$\,cm$^{-3}$ at the end of the prestellar phase and during the protostellar phase. Figure~\ref{fig-nhav} shows the resulting evolution of $n_{H,0}$ and $n_{H,0.55}$. WCCC probably cannot be reproduced for much higher densities because of a rapid freeze-out of carbon chains, preventing high gas-phase abundances. Thus, many simulations of WCCC cores consider densities of $\approx$$10^6$\,cm$^{-3}$ (see references in Section~\ref{intrd}).

As for a model that considers a single parcel that is not in the center of the core, irradiation is mostly regulated by $A_V$ and, thus, $N_H$, between the parcel and the closest edge of the core. Compared to radiation coming from the near edge (see $A_{V,0.55}$ in Table~\ref{tab-macr}), the radiation coming from the opposite side (through the center of the core) is reduced by 2.4\,mag at $t$=0 and 127\,mag in the protostellar stage at $t$=1.6--2.4\,Myr (cf. Figures \ref{fig-rad} and \ref{fig-nhav}). Thus, the far-side radiation affects chemistry in the initial stages and becomes increasingly unimportant in the advanced stages. Irradiation coming from the opposite side of the cloud was considered for a presumably more realistic chemical history for the considered parcel.

In the envelope, the gas parcel, for which we calculated chemistry, is located at a distance of 2640\,AU from the protostar. This value is higher than those known for WCCC cores; observations indicate radius of 500--1000\,AU for L1527 \citep{Sakai10} and 1500--2500\,AU for the protostar in the Lupus\,1 cloud \citep{Sakai09lupus}. However, as stated above, we do not aim to simulate the collapsing cloud core, instead limiting the model to providing an evolutionary history for a gas parcel that eventually produces WCCC in the protostellar envelope.

The irradiation intensity of our considered parcel of the cloud core is the key changing parameter. The reference model was assumed to be irradiated by a tenth of the standard ISRF photon and CR flux (see Section~\ref{mchm}), assuming that the core is partially shielded by its parent cloud complex. In other words, the standard intensity of these radiation types was reduced by a factor of 0.1. Hence the abbreviation of this model is ``1REF0.1''.

Figure~\ref{fig-phys} shows how the CR-induced ionization rate and temperature differs for the reference and irradiation models. In the prestellar stage, the temperature of the cloud was calculated as function of $A_V$. Gas temperature $T_{\rm gas}$ was calculated from the data of \citet{Hocuk16}, as in \citet{K17}:
\begin{eqnarray}
	\label{mmacr2}
	\hspace{-2cm}
	{\rm log}_{10}(T_{\rm gas}) = \nonumber \\
	3.019-2.215{\rm tanh}(2.135-{\rm log}_{10}(A_V))+ \nonumber \\
	1.542{\rm tanh}(0.8634-{\rm log}_{10}(A_V))^3+ \nonumber \\
	1.100{\rm tanh}(0.01823-{\rm log}_{10}(A_V))^5 \, .
\end{eqnarray}
Dust temperature $T_{\rm dust}$ was calculated with the approach of \citet{Hocuk17}, which, in addition to $A_V$, also depends on the ISRF ($G_0$), and thus differs for the reference and irradiation models. The value of $T_{\rm gas}$ was not allowed to be lower than that of $T_{\rm dust}$.

For the protostellar stage, $T_{\rm gas}$ and $T_{\rm dust}$ were calculated with the $T_2$ profile of \citet{Garrod06} and \citet{Garrod08}. We adopted the long and slow heating time-scale, relevant for low-mass protostars, where a temperature of 200\,K is reached in 1\,Myr. From $t=1.62$ to 2.45\,Myr, the temperature rises from $\approx10$\,K at the end of the prestellar stage to 142\,K (the maximum temperature in the model) at the end of the protostellar envelope stage. The simulations end shortly after sublimation of all water ice. Higher temperatures are irrelevant for this investigation, which focuses on the lukewarm WCCC cores.

\subsection{Chemical model} \label{mchm}

\textsc{Alchemic-Venta} is a modified kinetic rate-equation model, based on the ALCHEMIC code \citep{Semenov10}. We employ the UMIST database for astrochemistry 2012 (UDfA12) as the gas-phase reaction network \citep{McElroy13}. This network includes ionization and dissociation by CR protons, CR-induced photons, and the ISRF, as well as binary reactions of neutral and ionic species, including recombination of cations. The UDfA12 network makes our model different from previous WCCC simulations, which employ networks based on the OSU database \citep{Aikawa08,Aikawa20,Harada08,Hassel08,Hassel11,Wang19}. Compared to OSU, UDfA12 has recalculated photoreaction rates for the interstellar (Draine field) and CR-induced photons. These differences may cause some disagreement with the results of the previous simulations. To prevent the accumulation of large amounts of hydrogen at low temperatures, the H$_2$ contact desorption was employed \citep{Hincelin15}.

%
% Table 2
\begin{table}
\begin{center}
\caption{Initial abundances of chemical species relative to hydrogen.}
\label{tab-abds}
\begin{tabular}{lcc}
\hline\hline\\
Species & Abundance & Reference \\
\hline
H$_2$ & 0.500 &  \\
He & 0.0900 & 1 \\
C$^+$ & $7.30\times10^{-5}$ & 1 \\
N & $2.14\times10^{-5}$ & 1 \\
O & $1.76\times10^{-4}$ & 1 \\
F & $6.68\times10^{-9}$ & 1 \\
Na$^+$ & $2.25\times10^{-9}$ & 2 \\
Mg$^+$ & $1.09\times10^{-8}$ & 2 \\
Si$^+$ & $9.74\times10^{-9}$ & 2 \\
P$^+$ & $2.16\times10^{-10}$ & 2 \\
S$^+$ & $9.14\times10^{-8}$ & 2 \\
Cl & $1.00\times10^{-9}$ & 1 \\
Fe$^+$ & $2.74\times10^{-9}$ & 2 \\
\hline
\end{tabular} \\
References. 1: \citet{Wakelam08}; 2: \citet{Aikawa99}
\end{center}
\end{table}
Table~\ref{tab-abds} shows the initial elemental abundances of chemical species. We employ low-metal abundances, which result in an ice mantle thickness of about 85 monolayers (MLs), when total freeze-out occurs. The simulation takes a few $10^4$ years for the chemistry to relax to quasi-equilibrium. Real chemical equilibrium is never attained for any species because of the changing physical conditions.

The ISRF was assumed to correspond to a photon flux of $1.7\times10^8\,\rm cm^{-2}\,s^{-1}$ (note, the radiation comes from two opposite sides, as shown in Figure~\ref{fig-rad}). Self-shielding of H$_2$, and self- and mutual shielding (by H$_2$) of CO and N$_2$ from the ISRF photons was considered with the help of the tabulated data by \citet{Lee96} and \citet{Li13}. Shielding of icy surface molecules was not considered because their UV absorption bands are shifted in relation to those of gaseous species. The model includes desorption of products from photodissociation (ISRF and CR-induced) of icy species on the outer surface layer of icy grains, according to the general approach outlined in Appendix~C of \citet{K18mn}.

UDfA12 has more than sufficient chemical diversity for considering both carbon-chain species and COMs for the aims of this study. The network for surface reactions was adapted from the COMs network by \citet[][based on the OSU database]{Garrod08} with changes from \citet{Laas11} and \citet{K15apj2}. This network includes more organic species than the UDfA12 network, thus the COMs surface network was reduced accordingly. The surface binary reaction rate was adjusted by reaction-diffusion competition \citep{Garrod11}. Photodissociation rates of icy species were adopted from UDfA12 and thus are updated and considerably different from the original COMs-OSU network data. Following \citet{K18mn}, surface photoreaction rates were reduced by a factor of 0.3 in relation to their gas-phase counterparts. Reactions on the outer surface of the icy grain and within the bulk-ice were included.

The icy mantle of molecules adsorbed onto grain surfaces was described with the multi-layer approach \citep[see][]{K15apj1,Furuya17}. In addition to the 1--2\,ML thick surface layer, three bulk-ice layers with active chemistry were considered. This is sufficient to resolve polar (H$_2$O-dominated) and non-polar (CO-dominated) components of evolved icy mantles \citep{Sandford88}. For binary surface reactions, the ratio between molecule binding (or diffusion) energy $E_{b,s}$ and desorption energy $E_D$ was taken to be 0.50 \citep{Garrod08}. Reactions can occur either via molecule hopping or tunneling, whichever is faster \citep{Garrod13}. The ratio between the absorption energy of bulk-ice species $E_B$ and $E_D$ was taken to be $E_B/E_D=2$ and, consequently, the binding energy of mantle species is $E_{b,m}=1.0E_D$. $E_{b,m}$ was used to calculate the rate of bulk ice reactions with the approach outlined in \citet{K15apj1} and the rate of diffusion between the four ice mantle layers \citep{KK19}.

The model considers olivine grains with radius $a=0.1\,\mu$m, density of 3\,g\,cm$^{-3}$ constituting 1\,\% of cloud mass. When adsorption occurs, the grains become covered with molecules occupying an assumed average volume of a cube with a size of $3.2\times10^{-8}$\,cm. Such a molecule volume corresponds to water ice with a density of 0.9\,g\,cm$^{-3}$. The surface density of adsorption sites is $9.8\times10^{14}$\,\,cm$^{-2}$. The number of adsorption sites in successively adsorbed MLs increases in accordance with the increasing grain size. All neutral gaseous species were allowed to become adsorbed on grain surfaces. The molecule sticking coefficient was taken to be unity, with the exception of hydrogen, for which the coefficient was calculated following \citet{Thi10}.

The model considers several desorption mechanisms of icy surface species. Desorption by ISRF photons regulates the onset of ice accumulation at low $A_V$ values. CR-induced photons ensure that some desorption occurs even at high extinctions. For CH$_4$, CO, CO$_2$, CH$_3$OH, N$_2$, NH$_3$, and H$_2$O, experimental photodesorption yields $Y_{\rm ph}$ for both processes were adopted \citep[see Table~3 of][and references therein]{K18mn}. For all other icy species, $Y_{\rm ph}$ was taken to be 0.001. Another important mechanism is simple thermal evaporation. It is the principal desorption mechanism in the second stage of the model, which simulates heating of a circumstellar envelope. Reactive (chemical) desorption was also included, simulating the ejection of products of exothermic surface reactions. The desorption probability was calculated with the Rice-Ramsperger-Kessel theory, as explained by \citet{Garrod06f,Garrod07}. The efficiency parameter $a$ was taken to be 0.03.

The above desorption processes were supplemented by CR-induced whole-grain heating, resulting in CRD. This desorption mechanism was considered in particular detail, thanks to improvements made for the \textsc{Alchemic-Venta} model made in our previous study \citep{KK19}. In addition, recent advances in the understanding of evaporative cooling of icy grains have led to an improved CRD rate calculation. The CRD mechanism is described in detail in Appendix~\ref{app-crd}.

The rate of hydrogen ionization by CR protons $\zeta$ was calculated following \citet{Ivlev15p}, model ``High''. This approach gives $N_H$-dependent curve of $\zeta$ (Figure~\ref{fig-phys}). The flux of CR-induced photons, used for CR-induced photodesorption, was made proportional to $\zeta$ \citep{Cecchi92,KK19}.

\section{Results} \label{rslt}

%
% Table 3
\begin{table*}
\begin{center}
\caption{List of considered models.}
\label{tab-mdls}
\small
\begin{tabular}{cllccl}
\hline\hline\\
& & \multicolumn{3}{r}{Flux multiplier} &  \\
No. & Name (type) & Abbrev. & ISRF & CR & Description \\
\hline
1 & Reference & 1REF0.1 & 0.1 & 0.1 & Cloud core, shielded by the parent cloud complex \\
2 & Irradiation & 2IRR1.0 & 1.0 & 1.0 & Cloud core exposed to full irradiation from outside \\
3 & ISRF irradiation & 3ISRF1.0 & 1.0 & 0.1 & Core exposed to interstellar photons only \\
4 & CR irradiation & 4CR1.0 & 0.1 & 1.0 & Core exposed to CRs \\
5 & $\zeta$ test & 5ZETA1.0 & 0.1 & 0.1 & Model ``1REF0.1'' with $\zeta$ and CR-induced photon flux \\
 & & & & & increased by a factor of 10 \\
 & & & & & (i.e., the same $\zeta$ as in ``2IRR1.0'' and ``4CR1.0'' models) \\
6 & CRD test & 6CRD1.0 & 0.1 & 0.1 & Model ``1REF0.1'' with CRD rate increased by a factor of 10  \\
 & & & & & (i.e., the same CRD rate as in the ``2IRR1.0'' ``4CR1.0'' models) \\
\hline
\end{tabular}
\end{center}
\end{table*}
As discussed in Section~\ref{mmacr}, two main astrochemical simulations of a star-forming cloud core were run for this study: the reference model and the irradiation model. The reference cloud core (Model 1REF0.1 in Table~\ref{tab-mdls}) was assumed to be shielded by its parent cloud complex and irradiated by ISRF and CRs at 0.1 of their standard intensity. The irradiation model 2IRR1.0 considers a core subjected to standard radiation fields without shielding.

In interpreting the modeling results, we use the notion by \citet{Garrod08} that the temporal chemical evolution of a star-forming core (see Figure~\ref{fig-chem}) also qualitatively represents its 1D spatial structure, with less evolved regions located near the outer rim of the core.

WCCC regions are defined by their density ($\approx$$10^5-10^7$\,cm$^{-3}$) and temperature ($\approx$13--35\,K). Figures \ref{fig-cc1}, \ref{fig-cc2} and other plots focusing on carbon chains show $t$ interval of 0.8--2.0\,Myr, encompassing the major features in the evolution of the carbon- chain abundances, up to a temperature of 40\,K. Given the variety of the observed WCCC molecules, each figure displays different species, whenever reasonable. The calculated abundances were compared to those observed in L1527, which is the first detected and probably the most studied WCCC core \citep[e.g.][]{Yoshida19}. Compared to other WCCC sources, L1527 has relatively high abundances of carbon chains. However, these abundances are not outside the realm of those observed in other cores, thus L1527 may serve as an example of a `classic', pronounced WCCC source.

\subsection{Comparison of reference and irradiation models} \label{rsal}
%
% Figure 4
\begin{figure*} [htb!]
\vspace{-1.0cm}
\hspace{-2.0cm}
%\plotone{fig-chem.eps}
\includegraphics[width=23cm]{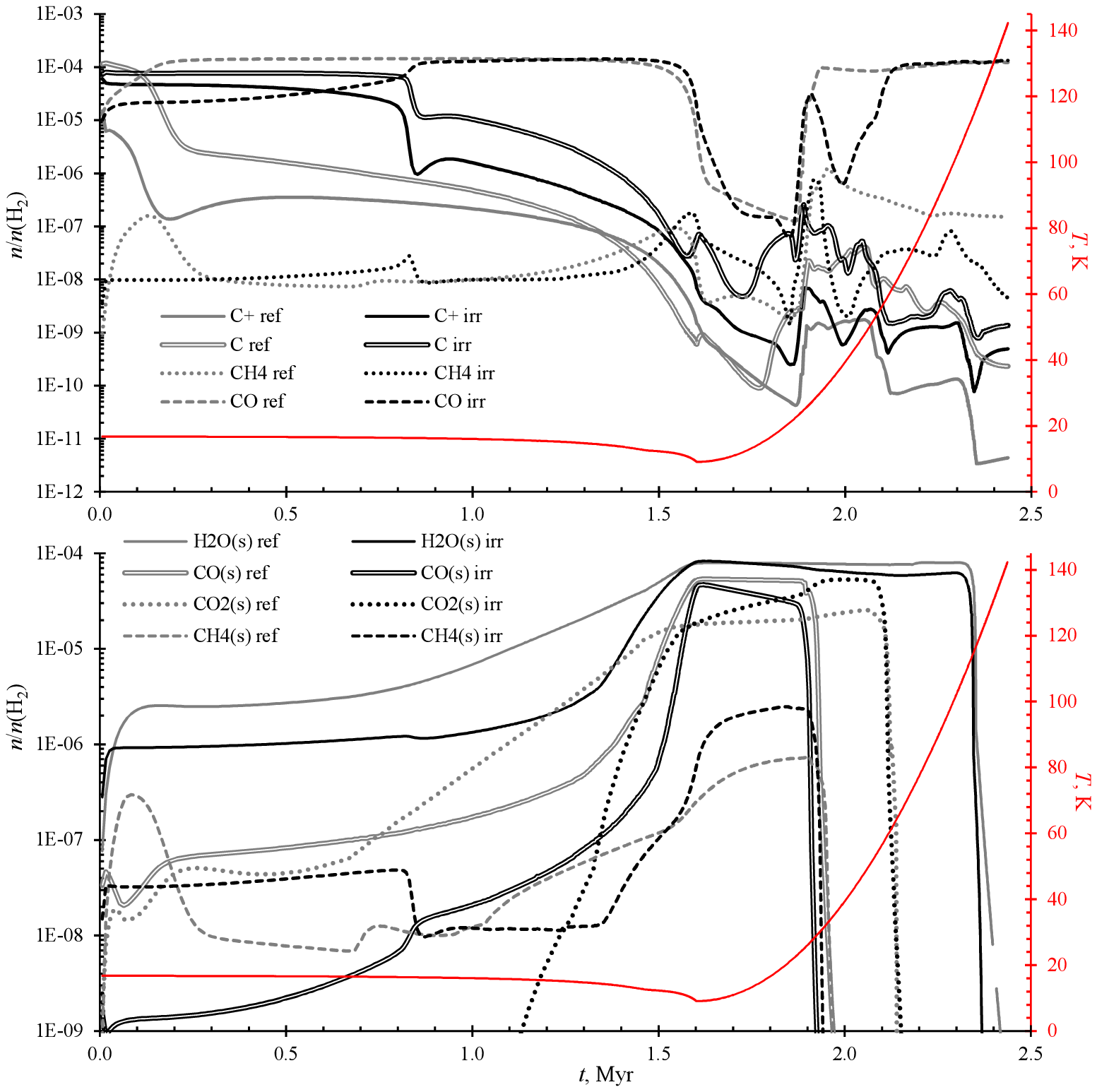}
\vspace{-13.5cm}
\caption{Calculated abundances relative to molecular hydrogen for the reference 1REF0.1 (``ref'' in the figure) and irradiation 2IRR1.0 (``irr'') models for important carbon gaseous species (top) and major icy molecules (bottom). The red curve indicates the temperature of the gas parcel in the 2IRR1.0 model, measured with the right-hand vertical axis. The abbreviation (s) indicates solid surface icy species.
 \label{fig-chem}}
\end{figure*}

Figure~\ref{fig-chem} shows the evolution of the abundance for a few of the most important gaseous species that affect production of carbon chains. For the irradiation model, CO becomes the main carbon species at $t$=0.81\,Myr.

After (or behind) the CO photodissociation layer, the irradiation model maintains a higher abundance of atomic carbon, which benefits the synthesis of methane and carbon chains. WCCC has been associated with gaseous methane, evaporated from icy grains in the circumstellar envelope (Section~\ref{intrd}). In the irradiation model, methane ice abundance increases by a factor of up to four, relative to the reference model. CRD prevents more effective accumulation of CH$_4$ ice in the irradiation model 2IRR1.0, where CRD efficiency is ten times higher than in 2REF0.1 (Table~\ref{tab-mdls}).

%
% Figure 5
\begin{figure*} [htb!]
\vspace{-1.0cm}
\hspace{-2.0cm}
%\plotone{fig-cc1.eps}
\includegraphics[width=23cm]{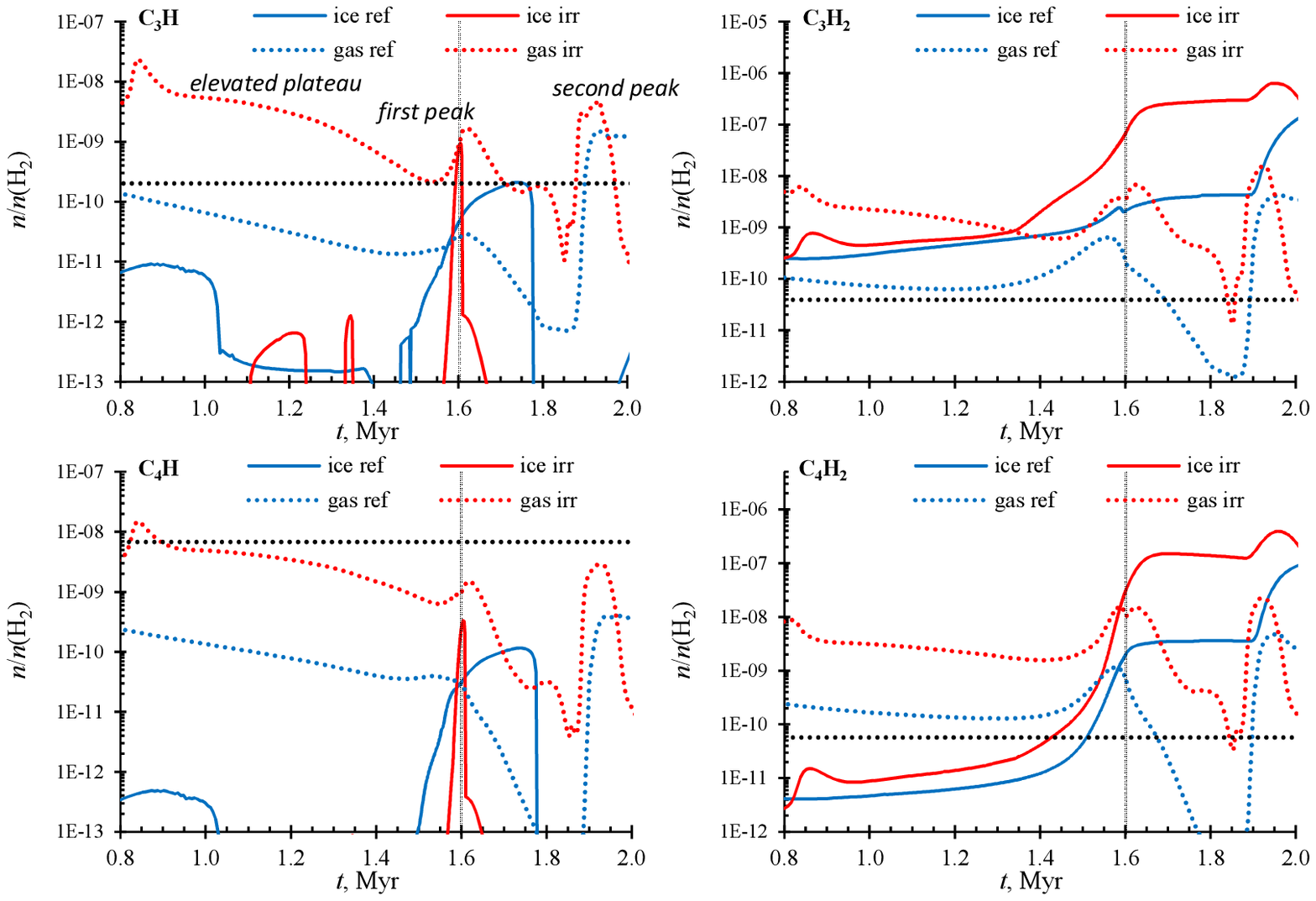}
\vspace{-18.5cm}
\caption{Calculated abundances relative to molecular hydrogen for typical carbon-chain species in reference (1REF0.1, ``ref'') and irradiation (2IRR1.0, ``irr'') models. The vertical gray line indicates the protostar formation time, while the dotted horizontal line indicates abundances observed in L1527 \citep{Sakai08,Araki17}. The model does not discern between c-C$_3$H, c-C$_3$H$_2$, and l-C$_3$H, l-C$_3$H$_2$, i.e., chain and cyclic isomers are treated together as single species with a single molecular formula.
\label{fig-cc1}}
\end{figure*}

In the time interval of interest, the 2IRR1.0 model produces abundances of carbon chains that are, typically, from a few to a few hundred times higher than those of the 1REF0.1 model. This result generally confirms that the WCCC phenomenon can be caused by locally increased irradiation or exposure of a star-forming core to the ISM. Figure~\ref{fig-cc1} shows that, in the irradiation model, a broad plateau of elevated carbon-chain abundances is followed by two abundance peaks in the 1.5--2.0\,Myr interval. The plateau starts with a small peak at $\approx$0.85\,Myr, when self-shielding prevents the destruction of CO, thus reducing the overall ionization level in the cloud, which promotes the formation of multi-atomic species. The small peak occurs at densities of $\approx$3000\,cm$^{-3}$ and is not counted as relevant to WCCC observations that sample gas with $n_H\geq10^5$\,cm$^{-3}$.

The first (or the outer) peak occurs at $\approx$1.6\,Myr, at a stage when the evolved core rapidly collapses, increasing its density and forming a protostar in the center. Any remaining free carbon atoms in the gas phase form molecules that soon freeze-out onto the grains. In the irradiated core, there is more CO still in the gas because of a more effective CRD. At the moment of starbirth ($t$=1.602\,Myr), the abundance of gaseous CO, $n_{\rm CO}$ is higher by a factor of 3.1 in the irradiation model, reaching $1.8\times10^{-5}$ relative to H$_2$. The abundance of gaseous C atoms is 94 times higher in the irradiation model, where it is $5.6\times10^{-8}$ relative to H$_2$, cf. Figure~\ref{fig-chem}.  More abundant C benefits to the formation of carbon chains. The abundance of atomic O in the 2IRR1.0 model is increased only by a factor of 2.7. Carbon-chain formation triggered by oxygen depletion was proposed already by \citet{Cordiner12-}.

After the first peak, the relative abundance of gaseous carbon chains drops by one or more orders of magnitude. During this intermediate time ($\approx$1.66--1.86\,Myr), the irradiation model retains carbon chain abundances that are 1--3 orders higher than those in the reference model. This is because of a higher abundance on the icy outer surface of the circumstellar grains, where the carbon chains are exposed to non-thermal desorption mechanisms. For the latter, CRD and CR-induced photons are crucial. The latter causes photodesorption and dissociation of molecules, which, in turn, results in reactive desorption. In the irradiation model, chains are depleted later, and thus have a few orders of magnitude higher abundance in the outer icy surface layer. Moreover, the irradiation model also has ten times higher flux of CR-induced photons.

The second peak at $t$$\approx$1.92\,Myr occurs concurrently with methane ice sublimation, and arises thanks to gas-phase processing of CH$_4$. It can be regarded as the ``true'' WCCC phenomenon in the model. Interestingly, the gas-phase peak abundance for methane is higher in the 1REF0.1 model compared to 2IRR1.0 ($1.3\times10^{-6}$ versus $7.5\times10^{-7}$, relative to H$_2$, respectively, cf. Figure~\ref{fig-chem}). This is because the CRs and CR-induced photons in the irradiation model rapidly destroy CH$_4$, transferring its carbon to the chains.

\subsection{Comparison with observations} \label{robs}
%
% Figure 6
\begin{figure*} [htb!]
\vspace{-1.0cm}
\hspace{-2.0cm}
%\plotone{fig-cc2.eps}
\includegraphics[width=23cm]{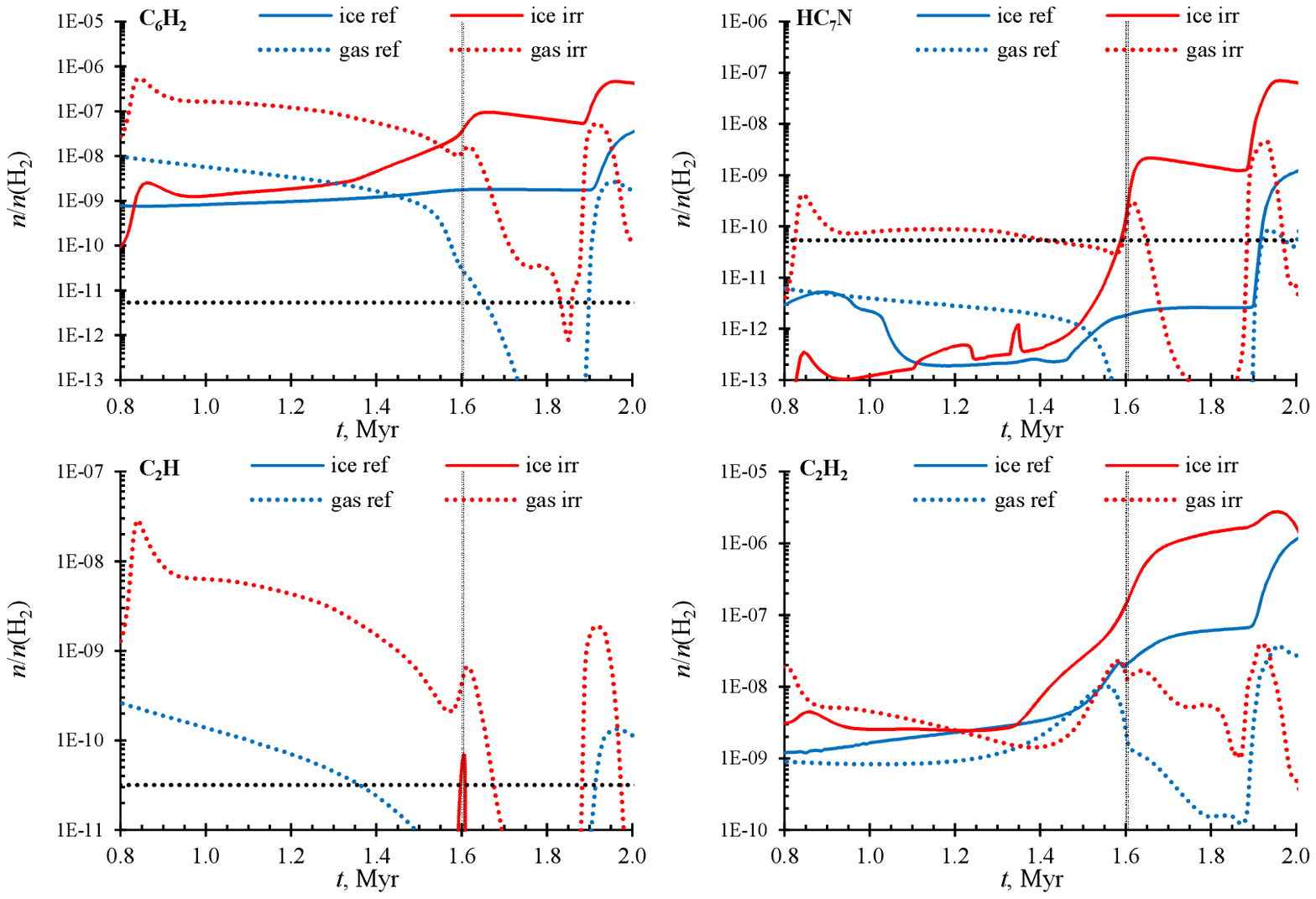}
\vspace{-18.5cm}
\caption{Calculated abundances relative to molecular hydrogen for selected carbon-chain species, along with their observed abundances in L1527, when available \citep{Sakai07,Sakai08}. Details as in Figure~\ref{fig-cc1}.
 \label{fig-cc2}}
\end{figure*}
In accordance with the aims of this study, we present only a qualitative comparison of calculation results with observational data. A detailed quantitative comparison requires considering a full line of sight, for which at least one-dimensional chemical model is necessary. Such a 1D study will likely be performed in the future.

Observations show coexisting central peak and an extended component for carbon chains in WCCC cores, with no clear boundary between them \citep{Sakai10, Araki17}. Such a picture can be consistent with the results of the 2IRR1.0 model, if we assume that the elevated plateau, the first, and the second peaks are intermixed along the complex line-of-sight through a real star-forming core. In the elevated plateau phase, gas has a notably lower density than $n_H$ in observations ($\approx$$10^4$ versus $\approx$$10^6$\,cm$^{-3}$), and thus might not be relevant for WCCC. The elevated plateau and the first peak can be considered as ``remnant'', components, occurring in outer regions not yet significantly affected by protostellar activity, while the second peak may explain the centrally concentrated WCCC component, where methane is sublimated from ices.

In this qualitative study that does not consider a full line of sight, a ``successful'' reproduction of observational data can be assumed when the calculated abundance of a species exceeds the observed value. Figures \ref{fig-cc1}--\ref{fig-cc2} show that for most carbon-chain species at least one of the peaks exceeds the observed abundance. In observations, C$_3$H$_2$ has been detected only in the central component, apparently corresponding to the second peak in the model. On the other hand, C$_2$H and C$_4$H have an extended structure \citep{Sakai10}, corresponding to the elevated plateau and the first peak in the 2IRR1.0 model. Such a behavior is approximately reproduced in the 2IRR1.0 model results, where the extended component for C$_3$H$_2$ is relatively much weaker than that of C$_2$H and C$_4$H$_2$, while the second peak is strong for all three molecules (cf. Figures \ref{fig-cc1} and \ref{fig-cc2}).

Another issue is the presence of COMs in WCCC cores. The correlation between the abundances of carbon chains and COMs is complex \citep{Graninger16,Lindberg16,Higuchi18} but it is expected that a ``classical'' WCCC source, such as the one we are attempting to simulate with the 2IRR1.0 model, is abundant in carbon chains and low on COMs. Surveys comparing the abundances of only C$_2$H and C$_4$H to those of CH$_3$OH have been done by \citet{Graninger16} and \citet{Lindberg16}.

%
% Figure 7
\begin{figure*} [htb!]
\vspace{-1.0cm}
\hspace{-2.0cm}
%\plotone{fig-com.eps}
\includegraphics[width=23cm]{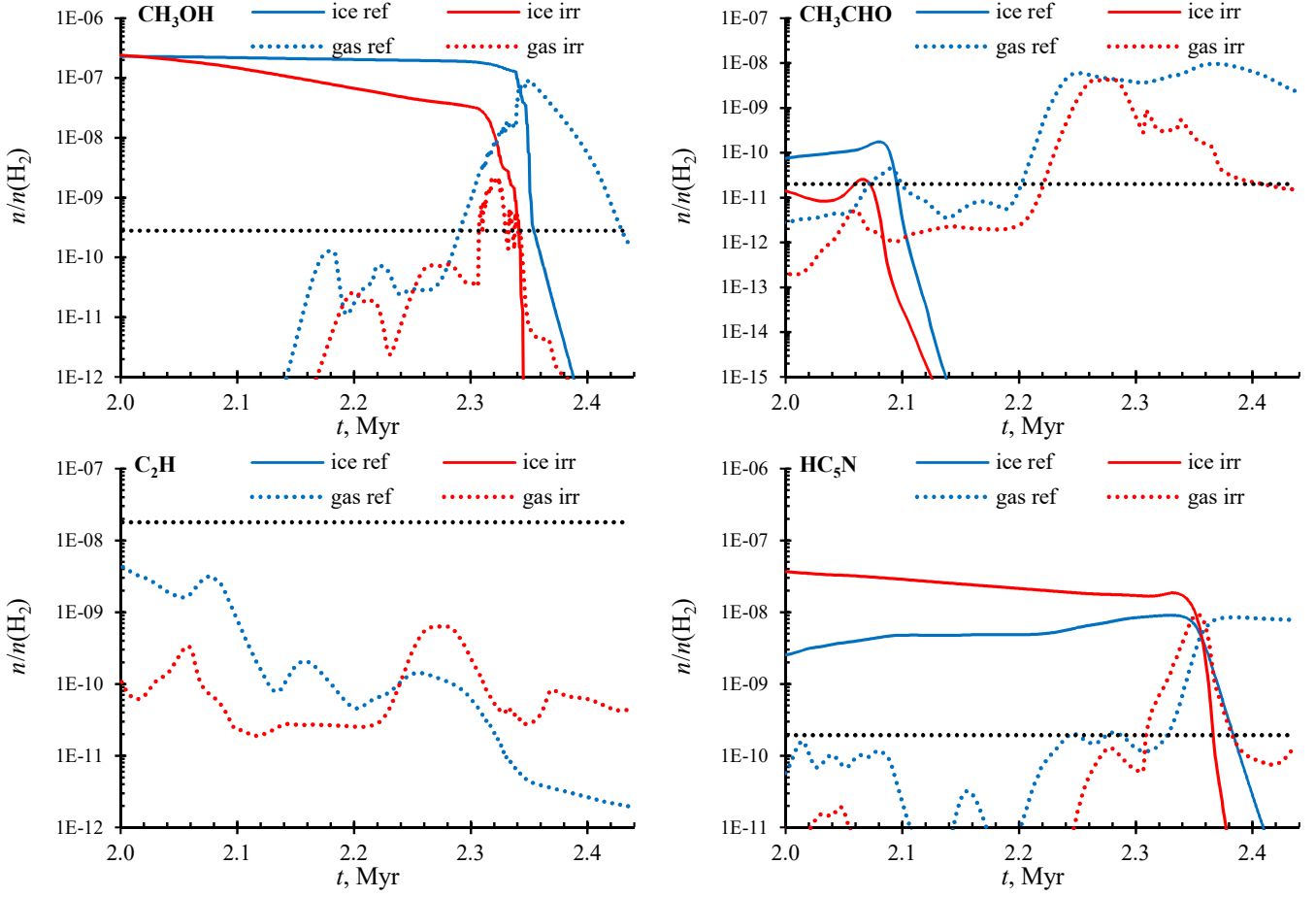}
\vspace{-18.5cm}
\caption{Calculated abundances relative to molecular hydrogen for examples of COMs and carbon chains during the late stage of the 1REF0.1 and 2IRR1.0 models. Details as in Figure~\ref{fig-cc1}, observed abundances shown for L1527 \citep{Sakai08,Yoshida19}.
 \label{fig-com}}
\end{figure*}

Figure~\ref{fig-com} shows a comparison of abundances for typical COMs -- methanol and acetaldehyde to those of example carbon chains in the latter stages of the model. At the temperatures relevant for hot corinos ($T_{\rm gas}>100$\,K, $t>2.3$\,Myr), the COMs have significantly lower abundances in the irradiation model. For carbon chains, there no clear trend, the carbon-chain abundances of 2IRR1.0 can be higher, comparable, or lower than those in the 1REF0.1 model, depending on species. This is because the rate of CR-induced photodissociation for CO is temperature-dependent in the UDfA12 network, which maintains a relatively high abundance of atomic C late in the simulation, resulting in high abundances for some carbon chains late in the simulations.

Summarizing, the calculated abundances of carbon chains in the temperature region relevant to WCCC cores are considerably higher in the irradiation 2IRR1.0 model, when compared to the reference 1REF0.1 model. The irradiation model  also apparently reproduces the few chemical features known about ``classical'' WCCC cores -- extended and central components of carbon chains, and a lower abundance (relative to the reference model) of COMs at their respective temperatures of $\geq100$\,K in the circumstellar envelope. Quantification of these features would require	 a more adequate macrophysical model.
%uz conclusions?

%
% Figure 8
\begin{figure*} [htb!]
\vspace{-1.0cm}
\hspace{-2.0cm}
%\plotone{fig-1d.eps}
\includegraphics[width=23cm]{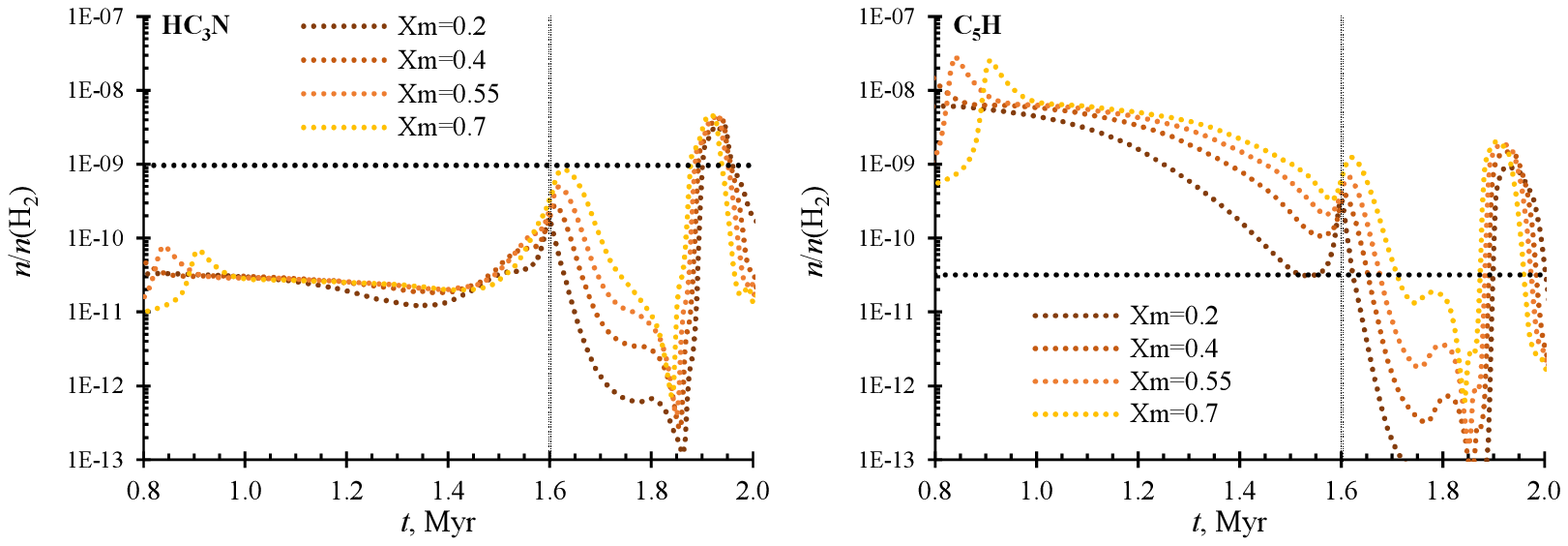}
\vspace{-24.5cm}
\caption{Comparison of calculated example carbon-chain gas-phase abundances in the 2IRR1.0 models at four different relative mass coordinates $X_m$. Details as in Figure~\ref{fig-cc1}; observed abundances in L1527 from \citet{Sakai08} and \citet{Sakai09deut}.
 \label{fig-1d}}
\end{figure*}
Figure~\ref{fig-1d} shows calculation results for the 2IRR1.0 model at relative mass coordinates 0.2, 0.4, 0.55, and 0.7 (in the direction from core center to its outer edge). These $X_m$ values correspond to radius from 1600 to 3000\,AU, although we emphasize again that this simple model does not adequately represent the structure of the star-forming core. The simulations at the four $X_m$ values all have similar temperature curves, although in a real star-forming core it can be expected that circumstellar envelope shells closer to the star are heated faster and reach higher temperatures. However, a qualitative trend can be obtained from the data in Figure~\ref{fig-1d} -- the carbon chains reach higher abundances in the outer layers of the circumstellar envelope. This is primarily because of a lower density in the model ($n_{H,0.7}=8\times10^5$, while $n_{H,0.2}=2\times10^6$), which does not allow rapid freeze-out of carbon chains that have just been formed in the gas phase.

\subsection{The causes of WCCC} \label{rmech}
%
% Figure 9
\begin{figure*} [htb!]
\vspace{-1.0cm}
\hspace{-2.0cm}
%\plotone{fig-icr.eps}
\includegraphics[width=23cm]{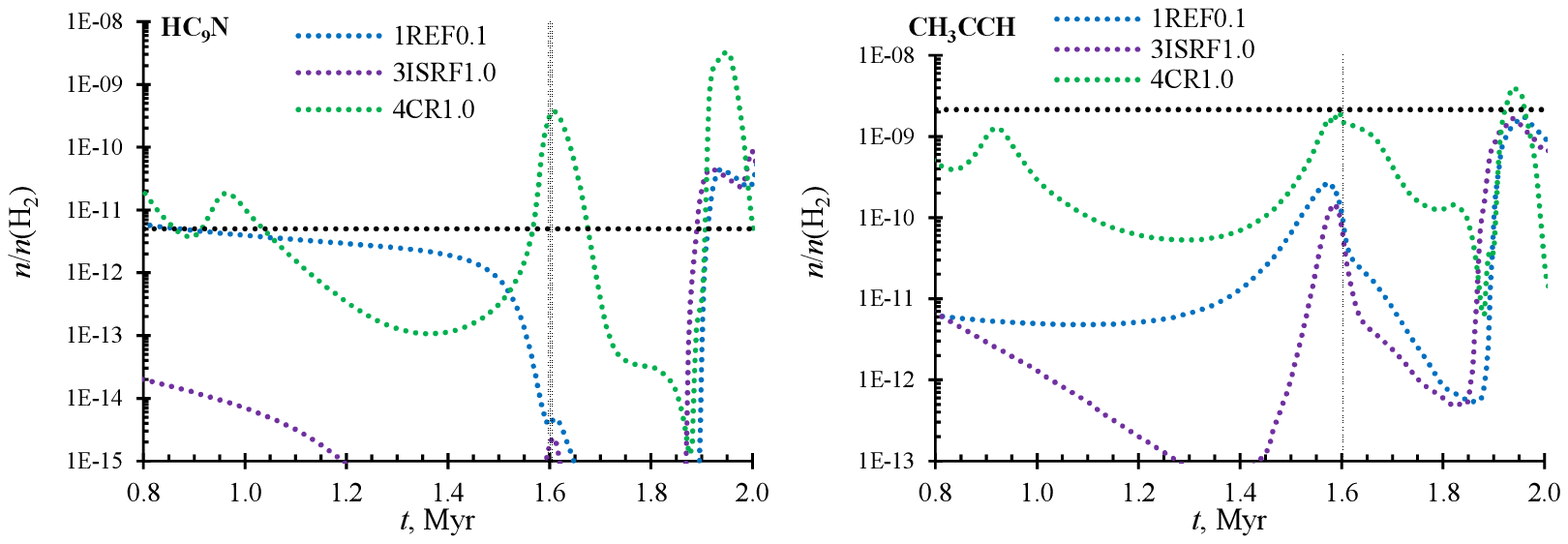}
\vspace{-24.5cm}
\caption{Comparison of calculated gas-phase abundances of examples of carbon-chains in the 1REF0.1 (``ref''), 3ISRF1.0 (``isrf''), and 4CR1.0 (``cr'') models. Observed abundances in L1527 from \citet{Sakai08}. The vertical gray line indicates the protostar formation time.
 \label{fig-icr}}
\end{figure*}
To determine the cause of the elevated carbon chain abundances in the irradiation model, two additional simulations were run. Model 3ISRF1.0 has interstellar photon intensity at its standard value (1.0) as in the irradiation model, but has a reduced (0.1 of standard) CR intensity, similar to that in the reference model. On the other hand, model 4CR1.0 has standard intensity of CRs and a intensity for ISRF reduced by a factor of 0.1, as in the reference model (see Table~\ref{tab-mdls}).

Figure~\ref{fig-icr} shows that CRs in the 4CR1.0 model have a significantly greater effect than interstellar photons in producing high carbon-chain abundances, compared to those in the 1REF0.1 model. The first and the second peaks happen only due to CRs, while the ISRF and CRs can both contribute to the elevated abundance plateau before the peaks. The contribution of ISRF is reducing the dominance of CO as the dominant gas-phase carbon species, which allows for more carbon chains to form.

%
% Figure 10
\begin{figure*} [htb!]
\vspace{-1.0cm}
\hspace{-2.0cm}
%\plotone{fig-crcr.eps}
\includegraphics[width=23cm]{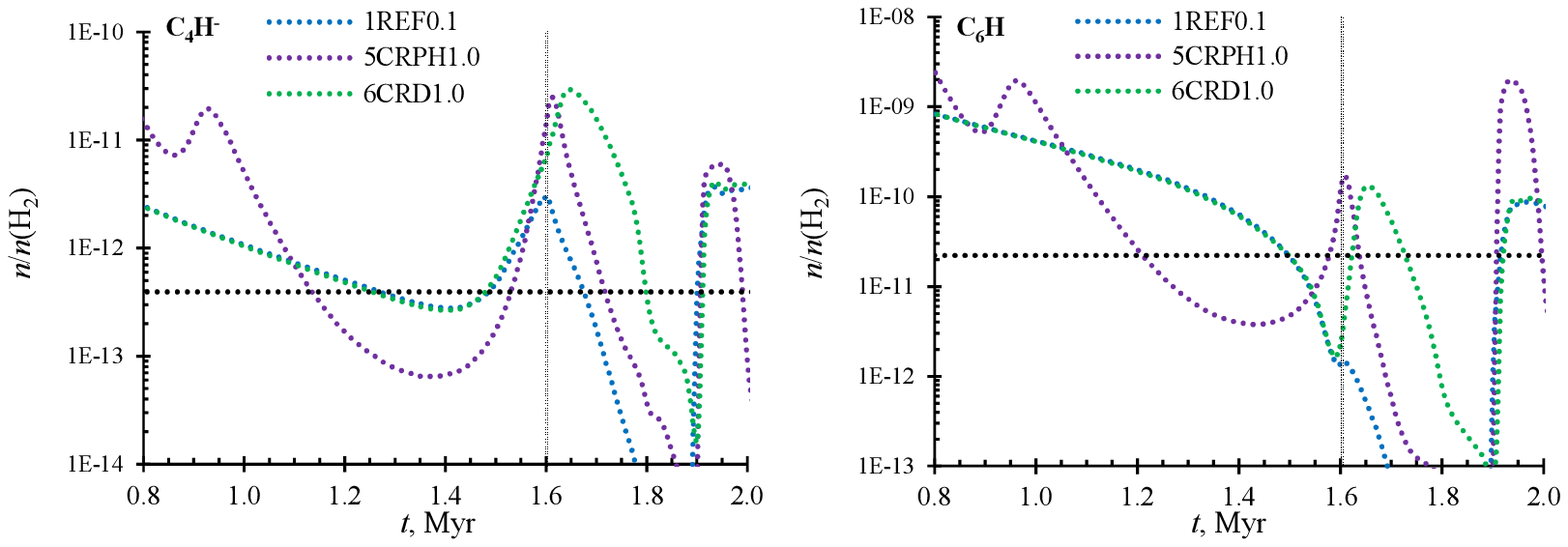}
\vspace{-24.5cm}
\caption{Comparison of calculated gas-phase abundances of examples of carbon-chains in the 1REF0.1, 5ZETA1.0, and 6CRD1.0 models. Observed abundances in L1527 from \citet{Sakai07} and \citet{Sakai08-}. The vertical gray line indicates the protostar formation time.}
 \label{fig-crcr}
\end{figure*}

The effects of CRs are twofold, arising from ionizing radiation (CR protons and their induced photons) and from grain heating by heavy CR nuclei, such as Si or Fe, which induces CRD (Appendix~\ref{app-crd}). Figure~\ref{fig-crcr} separates the effects from these two phenomena, comparing gas phase abundances for some carbon chains calculated with the reference model 1REF0.1, model 5ZETA1.0, and model 6CRD1.0 (see Table~\ref{tab-mdls}). Model 5ZETA1.0 is similar to the reference model, having the ISRF intensity and CRD frequency reduced by a factor of 0.1, compared to their standard values (Section~\ref{mchm}), while the CR-ionization rate $\zeta$, and, thus, also CR-induced photon intensity are at their standard values (flux multiplier is equal to 1.0). On the other hand, model 6CRD1.0 has its CRD frequency at standard value, while ISRF and $\zeta$ are reduced by a factor of 0.1 from their standard values.

The plots show that neither a higher CRD intensity, nor $\zeta$ are able to reproduce the elevated plateau. Both phenomena contribute to the first carbon-chain abundance peak, while only elevated $\zeta$ is able to reproduce the second peak, which is the actual WCCC region, not a remnant from the prestellar core.

From the above discussion, the following picture emerges. The early carbon-chain abundance plateau arises because of ionization and dissociation of molecules by the interstellar photons and CRs, which maintain higher abundances of C and C$^+$ later in the cloud evolution, allowing more carbon chains to form. The freeze-out of oxygen (in the form of water ice) in the late, dense stages of the prestellar core remove some destruction paths for carbon-chains, resulting in the first peak. CRs are essential for this peak, as the CRD process maintains elevated abundance of carbon in the gas phase in the form of the CO molecule, while CR-induced ionization detaches the carbon atoms from CO, allowing carbon chains to form. This same process allows also a few times more (compared to the reference model) of CH$_4$ ice to form on grain surfaces. Finally, the icy methane evaporates and is processed in the gas, producing the second peak. Because gas-phase chemistry is driven by CR-induced ionization, a sufficiently high value for $\zeta$ is essential for the second peak. The carbon chains synthesized in the gas are then dissociated by CR-induced photons, which makes the second peak a temporal phenomenon.

\section{Conclusions} \label{cncl}
%šeit paliku

Observations indicate that WCCC may arise in star-forming cores exposed to interstellar radiation. We have compared simulations of chemistry in a shielded core and in a core exposed to full irradiation by ISRF and CRs. The main findings are listed below.
\begin{itemize}
\item The irradiated core has carbon-chain gas-phase abundances higher by a factor of up to $10^3$, compared to the reference model. The exact increase depends on species and the evolutionary stage of the core and often results in molecules' calculated abundances being similar to or higher than observed abundances. This means that the irradiation model may better reproduce observations. The primary cause of the higher abundances is a higher ionizing radiation flux.
\item Three general features can be discerned in the evolution of carbon-chain abundances -- an elevated plateau, a first peak at star-formation time and a second, the ``true'' WCCC, peak at the temperature of methane evaporation. The plateau and the first peak can be considered as ``remnants'' from the prestellar stage.
\item The elevated plateau occurs at a relatively low density of $\approx$$10^4$\,cm$^{-3}$, which is lower by a factor of $\approx$100 than the densities associated with WCCC, and thus might be unobservable for some species. The plateau is caused by the ionization by the ISRF and CRs.
\item The first peak is a product of the combined effects of CR-induced ionization and CRD.
\item For the second ``true'' WCCC peak, only CR-induced ionization ($\zeta$) is essential. It induces an active gas-phase chemistry that is able to convert the carbon in evaporated methane molecules into carbon chains.
\item At lower densities, newly formed carbon chains freeze out slower and the WCCC phenomenon is more pronounced.
\item Unlike carbon chains, the abundances of COMs have no clear correlation with radiation, at least up to temperatures of $\approx$140\,K.
\end{itemize}

We conclude that WCCC is possible in star-forming cores that are sufficiently irradiated by CRs with $\zeta\gtrsim10^{-16}$\,s$^{-1}$. This value can be different for models employing chemical networks other than UDfA12. A star-forming core with a CR irradiation decreased by a factor of 0.1 (compared to the ``standard'' irradiation in the model) shows only weak WCCC features and is more abundant with COMs.

\acknowledgments

This research has been funded by ERDF postdoctoral grant No.~1.1.1.2/VIAA/I/16/194 ‘Chemical effects of cosmic ray induced heating of interstellar dust grains’ being implemented in Ventspils University of Applied Sciences. I am also grateful to Ventspils City Council for its support. This research has made use of NASA’s Astrophysics Data System. I thank the anonymous referees for the thorough review and many valuable comments that greatly improved the manuscript.

\software{ALCHEMIC \citep{Semenov10},  
          \textsc{Tcool} \citep{KK20} 
          }

\appendix
\section{Cosmic-ray induced desorption (CRD)}
\label{app-crd}

The current version of the \textsc{Alchemic-Venta} model takes into account the stochastic aspect of whole-grain heating by CRs by considering the ices in the heated grains as a separate physical ``warm" phase. The ambient grains are converted to hot grains with a rate $k_{\rm warm}$ and converted back (cool down) with a rate $k_{\rm cool}$. Such an approach is valid because even a relatively small parcel of the cloud core with a mass of, e.g., $10^{-4}\,M_\odot$ contains a huge number of heated grains (septillion or more), even with the low whole-grain heating rate employed by \citet{Hasegawa93}.

The warm ice phase differs from ambient icy grains with its temperature $T_{\rm CR}$, which is higher than the temperature $T_{\rm dust}$ of ambient grains. Chemical reactions, reactive desorption, inter-layer diffusion rates, and sublimation all occur at $T_{\rm CR}$. Photoprocessing was not considered for the warm phase because of its unimportance relative to icy molecule photoprocessing from ambient grains. On the other hand, chemical reaction (and thus, reactive desorption) and, especially, sublimation rates are strongly dependent on temperature. From all these processes, sublimation (i.e., CRD) is the one, which can significantly affect ice composition during the cloud core collapse stage \citep{KK19}.

Unlike our previous studies, in the present research we have the tools and data to acquire realistic values for $T_{\rm CR}$, $k_{\rm warm}$, and $k_{\rm cool}$. The first task is to choose a simple, yet justified and realistic, approach in calculating these parameters as functions of $N_H$ and amount of ices adsorbed on the 0.1\,$\mu$m grains. First, a single, constant $T_{\rm CR}$ can be employed. This is because the proportions of different sublimated species remain similar in 40--70\,K \citep[as shown by][]{KK20}, which is our $T_{\rm CR}$ range of interest (see below). Grain heating rate $k_{\rm warm}$ depends on CR intensity in the cloud, which is affected by gas column density \citep[e.g.][]{Strong98}, a changing parameter in the model. The cooling time $t_{\rm cool}$, which is inversely proportional to $k_{\rm cool}$ depends on the properties of the icy mantle -- thickness and amount of volatiles.

The heating of icy interstellar grains by CRs is a complex phenomenon, with grains carrying varying amounts of adsorbed ices being heated to various temperatures $T_{\rm CR}$ with various frequencies $f_T$. To obtain a single, characteristic $T_{\rm CR}$, we calculated the weighed average heating temperature from the complete grain heating energy spectra provided by \citet{K18aps}. The $T_{\rm CR}$ values were recalculated with the grain heat capacity $C$ derived by \citet{Leger85} because \citet{K18aps} partially used the simple Debye approach on $C$, which is inadequate for grain temperatures exceeding $\approx30$\,K \citep{KK20ii}. The weighed average $\bar T_{\rm CR}$ is different for grains with different ice layers: 64\,K for bare grains, 58\,K for grains with a 0.01\,$\mu$m thick ice mantle, 54\,K for grains with 0.02\,$\mu$m ice, and 50\,K for grains with 0.03\,$\mu$m ice. Interestingly, because of changes in CR spectra with increasing column densities \citep[see][]{Padovani09}, $\bar T_{\rm CR}$ for grains with 0.03\,$\mu$m ice decreases from 50\,K to 44\,K for $N_H$ of $2.2\times10^{22}$ to $5.1\times10^{23}$\,cm$^{-2}$, respectively. All these temperatures exceed the crucial 40\,K threshold, where the total energy loss becomes dominated by radiative, not evaporative cooling \citep{KK20}. We chose $\bar T_{\rm CR}$=54\,K as a compromise value. It corresponds to an ice mantle thickness representing partial freeze out of heavy molecules onto grains, i.e., representing a stage of ongoing accumulation of molecules in ices, when CRD most significantly affects ice composition.

%
% Figure 11(A1)
\begin{figure*} [htb!]
\vspace{-1.0cm}
%\hspace{-1.0cm}
%\plotone{fig-crd.eps}
\includegraphics[width=20cm]{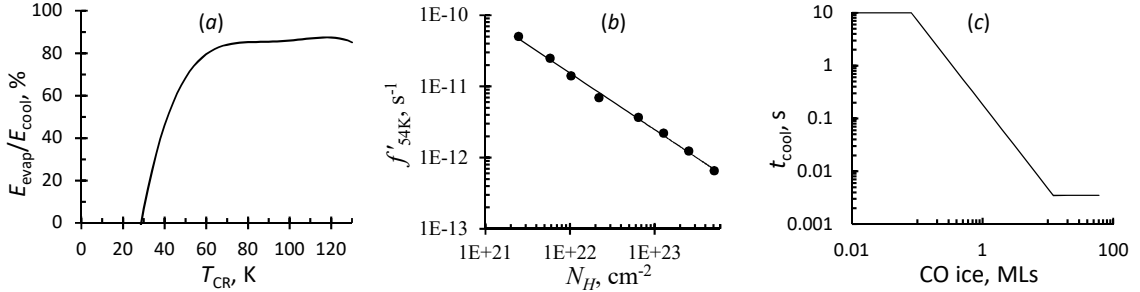}
\vspace{-23.0cm}
\caption{(\textit{a}): percentage of energy that a heated grain loses via sublimation of volatile species ($E_{\rm subl}$), relative to the total energy $E_{\rm cool}$ released during the cooling of the grain. (\textit{b}): CR-induced grain heating frequency to the assumed average heating temperature of 54\,K. The trendline drawn through the calculation points corresponds to Equation~\ref{mcrd2}. (\textit{c}): cooling time of a 0.1\,$\mu$m icy grain as a function of CO ice abundance on the grain, expressed in MLs, Equation~(\ref{mcrd3}). 
 \label{fig-crd}}
\end{figure*}
Knowing the chosen grain temperature allows calculating $f_T$. When a single $\bar T_{\rm CR}$ is used, it must represent all the energy, received by a grain from CRs, that is used for the sublimation of icy molecules. Naturally, this energy arises from impacts by CRs of various types, elevating the temperature of the icy grains to different values. The \textsc{Tcool} program, developed by \citet{KK20}, was used to calculate, how much energy $E_{\rm evap}$ goes away with sublimation of layered ices at different grain temperatures, compared to the total grain thermal energy lost during cooling $E_{\rm cool}$. Figure~\ref{fig-crd}\textit{a} shows these data graphically. The remainder of the energy goes away via radiative cooling, which in \textsc{Tcool} is calculated with the method of \citet{Cuppen06}.

The energy, used for sublimation at each temperature $T_{\rm CR}$, multiplied by the corresponding $f_T$ from the data of \citet{K18aps}, gives us the flux of CR energy $F_{E,\rm CR,evap}$ (eV\,s$^{-1}$) received by the grain and inducing sublimation of the interstellar ices. When all this energy is assumed to come from grain heating events with $\bar T_{\rm CR}$=54\,K, the corresponding frequency is
\begin{equation}
	\label{mcrd1}
	f^{'}_{54} = \frac{F_{E,\rm CR,evap}}{E_{\rm CR,54}X_{\rm subl,54}},
\end{equation}
where $E_{\rm CR,54}$ is the total energy imparted by a CR hit, heating a grain to 54\,K from the ambient $T_{\rm dust}$$\approx$10\,K and $X_{\rm subl,54}$ is the part of that energy used for sublimation. $E_{\rm CR,54}$ is equal to the energy lost by the grain, when it cools from 54\,K to $\approx10$\,K, which is about (2--4)$\times10^5$\,eV, depending on the thickness of the ice layer. $X_{\rm subl,54}$ was taken to be 0.74 from the data of \citet{KK20}. Finally, the rate coefficient for the transition of icy molecules into the warm ice phase is defined to be equal to $f^{'}_{54}$. Using the corresponding data, the CRD heating frequency numerically transforms into 
\begin{equation}
	\label{mcrd2}
	f^{'}_{54} = 4.20\times10^6 N_H^{-0.793}\,({\rm s}^{-1}),
\end{equation}
where $N_H$ is expressed in cm$^{-2}$. Figure~\ref{fig-crd}\textit{b} shows graphically the resulting heating frequency.

The reversal of warm ice species to the ambient ice phase occurs with a rate coefficient $k_{\rm cool}$. The established practice in astrochemistry is that $k_{\rm cool}$ is inversely proportional to the time of cooling $t_{\rm cool}$, specific for each given $T_{\rm CR}$. \citet{Hasegawa93} estimated $t_{\rm cool}$ as the characteristic evaporation time of the CO molecule, the primary volatile species in ices. Here, we retained the same general approach but defined $t_{\rm cool}$ as the time grain spends in $T_{\rm CR}$, during which the number of sublimated volatile molecules equals that sublimated during a realistic cooling of the same grain from $T_{\rm CR}$ to 10\,K. Only volatile species, such as CO and N$_2$ are able contribute to the cooling, before the grain cools down radiatively \citep{KK20}.

The thickness and composition of the icy mantle varies as the cloud core evolves. In the initial stages, the layer is thin and consists of mostly of non-volatiles, such as H$_2$O. In later stages, the proportion of CO ice reaches 20--40\,\% relative to water ice \citep{Whittet07}, while in the densest, coldest, and most shielded parts of the core this proportion may exceed 60\,\%, as shown by astrochemical models. The volatile-poor grains cool significantly longer, allowing more of their volatiles to escape, when compared to a cooling time that is similar to the CO evaporation time-scale.

The cooling times were calculated with the \textsc{Tcool} program. We considered a 0.1\,$\mu$m grain, covered by an icy mantle with a composition estimated from observations, when possible, or the results of \citet{KK19}. For example, a grain covered with 10\,MLs of ice containing 2\% of volatile molecules, relative to water ice, cools down from 54\,K to 10\,K releasing $2.2\times10^5$ molecules (CO, N$_2$, O$_2$, and CH$_4$), meaning that the grain lost $2.2\times10^4$\,eV via sublimation. The same amount of sublimated energy and molecules can be lost, when the grain is held at a constant temperature of 54\,K for 9.5\,s. For a grain with 60\,MLs of ice and 30\,\% volatiles, these numbers are $2.4\times10^6$ molecules, $2.4\times10^5$\,eV, and $5.2\times10^{-3}$\,s. For grains with a thick mantle and abundant (68\,\%) volatiles these numbers level out at $\approx3\times10^6$ molecules, $\approx3\times10^5$\,eV, and $\approx3.5\times10^{-3}$\,s. The exact values depend on the exact ice thickness, which affects grain heat capacity. When a sufficient amount of volatiles is present, the cooling time-scale remains fairly constant and about twice the characteristic CO evaporation time-scale, simply because the evaporative cooling requires the sublimation of molecules from two CO-dominated MLs.

From the data described above, the most precise approach to derive $t_{\rm cool}$ (s) is to express it as a function of the amount of volatiles in the icy mantle of the grain:
\begin{eqnarray}
	\label{mcrd3}
	\hspace{-2cm}
	t_{\rm cool} =  0.183\,{b_{\rm CO}}^{-1.59} \nonumber \\
	10\geq t_{\rm cool} \geq 3.5\times10^{-3} ,s ,
\end{eqnarray}
where $b_{CO}$ is abundance of CO ice, expressed in MLs. The result is shown in Figure~\ref{fig-crd}\textit{c}.

Summarizing the above, we have derived an $N_H$-dependent grain heating frequency $f_T$ and a $t_{\rm cool}$ value that is high for grains with low content of volatiles. This realistic approach effectively means that CRD is reasonably efficient only for volatile-poor grains at relatively low column densities, where it prevents an early accumulation \textit{en masse} of CO$_2$ and CO ices before their threshold $A_V$ values, which have been derived from observations \citep[4.3 and 6.7\,mag, respectively,][]{Whittet07}. CRD rapidly becomes inefficient when the first few layers of adsorbed icy CO appear, while the gas parcel is shifting to higher column densities.

For models that describe CRD of surface species with the simpler approach devised by \citet{Hasegawa93}, the ``duty cycle'' or time fraction spent by a grain in 54\,K is $f^{'}_{54}\times t_{\rm cool}$, and the CRD rate coefficient for species $i$ is
\begin{equation}
	\label{mcrd4}
	k_{\rm CRD}(i) = f^{'}_{54}t_{\rm cool}k_{\rm evap}(i,54\,{\rm K}) \,,
\end{equation}
where $k_{\rm evap}(i,54\,{\rm K})$ is the thermal desorption rate coefficient of species $i$ at 54\,K grain temperature.

\bibliography{wccc}{}

\begin{thebibliography}{}
\expandafter\ifx\csname natexlab\endcsname\relax\def\natexlab#1{#1}\fi
\providecommand{\url}[1]{\href{#1}{#1}}
\providecommand{\dodoi}[1]{doi:~\href{http://doi.org/#1}{\nolinkurl{#1}}}
\providecommand{\doeprint}[1]{\href{http://ascl.net/#1}{\nolinkurl{http://ascl.net/#1}}}
\providecommand{\doarXiv}[1]{\href{https://arxiv.org/abs/#1}{\nolinkurl{https://arxiv.org/abs/#1}}}

\bibitem[{{Ag{\'u}ndez} {et~al.}(2015){Ag{\'u}ndez}, {Cernicharo}, \&
  {Gu{\'e}lin}}]{Agundez15}
{Ag{\'u}ndez}, M., {Cernicharo}, J., \& {Gu{\'e}lin}, M. 2015, \aap, 577, L5,
  \dodoi{10.1051/0004-6361/201526317}

\bibitem[{{Aikawa} {et~al.}(2020){Aikawa}, {Furuya}, {Yamamoto}, \&
  {Sakai}}]{Aikawa20}
{Aikawa}, Y., {Furuya}, K., {Yamamoto}, S., \& {Sakai}, N. 2020, \apj, 897,
  110, \dodoi{10.3847/1538-4357/ab994a}

\bibitem[{{Aikawa} \& {Herbst}(1999)}]{Aikawa99}
{Aikawa}, Y., \& {Herbst}, E. 1999, \apj, 526, 314, \dodoi{10.1086/307973}

\bibitem[{{Aikawa} {et~al.}(2001){Aikawa}, {Ohashi}, {Inutsuka}, {Herbst}, \&
  {Takakuwa}}]{Aikawa01}
{Aikawa}, Y., {Ohashi}, N., {Inutsuka}, S.-i., {Herbst}, E., \& {Takakuwa}, S.
  2001, \apj, 552, 639, \dodoi{10.1086/320551}

\bibitem[{{Aikawa} {et~al.}(2008){Aikawa}, {Wakelam}, {Garrod}, \&
  {Herbst}}]{Aikawa08}
{Aikawa}, Y., {Wakelam}, V., {Garrod}, R.~T., \& {Herbst}, E. 2008, \apj, 674,
  984, \dodoi{10.1086/524096}

\bibitem[{{Araki} {et~al.}(2016){Araki}, {Takano}, {Sakai}, {Yamamoto},
  {Oyama}, {Kuze}, \& {Tsukiyama}}]{Araki16}
{Araki}, M., {Takano}, S., {Sakai}, N., {et~al.} 2016, \apj, 833, 291,
  \dodoi{10.3847/1538-4357/833/2/291}

\bibitem[{{Araki} {et~al.}(2017){Araki}, {Takano}, {Sakai}, {Yamamoto},
  {Oyama}, {Kuze}, \& {Tsukiyama}}]{Araki17}
---. 2017, \apj, 847, 51, \dodoi{10.3847/1538-4357/aa8637}

\bibitem[{{Aso} {et~al.}(2015){Aso}, {Ohashi}, {Saigo}, {Koyamatsu}, {Aikawa},
  {Hayashi}, {Machida}, {Saito}, {Takakuwa}, {Tomida}, {Tomisaka}, \&
  {Yen}}]{Aso15}
{Aso}, Y., {Ohashi}, N., {Saigo}, K., {et~al.} 2015, \apj, 812, 27,
  \dodoi{10.1088/0004-637X/812/1/27}

\bibitem[{{Bottinelli} {et~al.}(2004){Bottinelli}, {Ceccarelli}, {Lefloch},
  {Williams}, {Castets}, {Caux}, {Cazaux}, {Maret}, {Parise}, \&
  {Tielens}}]{Bottinelli04}
{Bottinelli}, S., {Ceccarelli}, C., {Lefloch}, B., {et~al.} 2004, \apj, 615,
  354, \dodoi{10.1086/423952}

\bibitem[{{Bouvier} {et~al.}(2020){Bouvier}, {L{\'o}pez-Sepulcre},
  {Ceccarelli}, {Kahane}, {Imai}, {Sakai}, {Yamamoto}, \&
  {Dagdigian}}]{Bouvier20}
{Bouvier}, M., {L{\'o}pez-Sepulcre}, A., {Ceccarelli}, C., {et~al.} 2020, \aap,
  636, A19, \dodoi{10.1051/0004-6361/201937164}

\bibitem[{{Cazaux} {et~al.}(2003){Cazaux}, {Tielens}, {Ceccarelli}, {Castets},
  {Wakelam}, {Caux}, {Parise}, \& {Teyssier}}]{Cazaux03}
{Cazaux}, S., {Tielens}, A.~G.~G.~M., {Ceccarelli}, C., {et~al.} 2003, \apjl,
  593, L51, \dodoi{10.1086/378038}

\bibitem[{{Cecchi-Pestellini} \& {Aiello}(1992)}]{Cecchi92}
{Cecchi-Pestellini}, C., \& {Aiello}, S. 1992, \mnras, 258, 125,
  \dodoi{10.1093/mnras/258.1.125}

\bibitem[{{Charnley} \& {Cordiner}(2010)}]{Charnley10}
{Charnley}, S.~B., \& {Cordiner}, M.~A. 2010, in American Astronomical Society
  Meeting Abstracts, Vol. 215, American Astronomical Society Meeting Abstracts
  \#215, 415.22

\bibitem[{{Cordiner} \& {Charnley}(2012)}]{Cordiner12-}
{Cordiner}, M.~A., \& {Charnley}, S.~B. 2012, \apj, 749, 120,
  \dodoi{10.1088/0004-637X/749/2/120}

\bibitem[{{Cordiner} {et~al.}(2011){Cordiner}, {Charnley}, {Buckle}, {Walsh},
  \& {Millar}}]{Cordiner11}
{Cordiner}, M.~A., {Charnley}, S.~B., {Buckle}, J.~V., {Walsh}, C., \&
  {Millar}, T.~J. 2011, \apjl, 730, L18, \dodoi{10.1088/2041-8205/730/2/L18}

\bibitem[{{Cordiner} {et~al.}(2012){Cordiner}, {Charnley}, {Wirstr{\"o}m}, \&
  {Smith}}]{Cordiner12cha}
{Cordiner}, M.~A., {Charnley}, S.~B., {Wirstr{\"o}m}, E.~S., \& {Smith}, R.~G.
  2012, \apj, 744, 131, \dodoi{10.1088/0004-637X/744/2/131}

\bibitem[{{Cuppen} {et~al.}(2006){Cuppen}, {Morata}, \& {Herbst}}]{Cuppen06}
{Cuppen}, H.~M., {Morata}, O., \& {Herbst}, E. 2006, \mnras, 367, 1757,
  \dodoi{10.1111/j.1365-2966.2006.10079.x}

\bibitem[{{Furuya} {et~al.}(2017){Furuya}, {Drozdovskaya}, {Visser}, {van
  Dishoeck}, {Walsh}, {Harsono}, {Hincelin}, \& {Taquet}}]{Furuya17}
{Furuya}, K., {Drozdovskaya}, M.~N., {Visser}, R., {et~al.} 2017, \aap, 599,
  A40, \dodoi{10.1051/0004-6361/201629269}

\bibitem[{{Garrod} {et~al.}(2006){Garrod}, {Park}, {Caselli}, \&
  {Herbst}}]{Garrod06f}
{Garrod}, R., {Park}, I.~H., {Caselli}, P., \& {Herbst}, E. 2006, Faraday
  Discussions, 133, 51, \dodoi{10.1039/b516202e}

\bibitem[{{Garrod}(2013)}]{Garrod13}
{Garrod}, R.~T. 2013, \apj, 765, 60, \dodoi{10.1088/0004-637X/765/1/60}

\bibitem[{{Garrod} {et~al.}(2017){Garrod}, {Belloche}, {M{\"u}ller}, \&
  {Menten}}]{Garrod17}
{Garrod}, R.~T., {Belloche}, A., {M{\"u}ller}, H.~S.~P., \& {Menten}, K.~M.
  2017, \aap, 601, A48, \dodoi{10.1051/0004-6361/201630254}

\bibitem[{{Garrod} \& {Herbst}(2006)}]{Garrod06}
{Garrod}, R.~T., \& {Herbst}, E. 2006, \aap, 457, 927,
  \dodoi{10.1051/0004-6361:20065560}

\bibitem[{{Garrod} \& {Pauly}(2011)}]{Garrod11}
{Garrod}, R.~T., \& {Pauly}, T. 2011, \apj, 735, 15,
  \dodoi{10.1088/0004-637X/735/1/15}

\bibitem[{{Garrod} {et~al.}(2007){Garrod}, {Wakelam}, \& {Herbst}}]{Garrod07}
{Garrod}, R.~T., {Wakelam}, V., \& {Herbst}, E. 2007, \aap, 467, 1103,
  \dodoi{10.1051/0004-6361:20066704}

\bibitem[{{Garrod} {et~al.}(2008){Garrod}, {Weaver}, \& {Herbst}}]{Garrod08}
{Garrod}, R.~T., {Weaver}, S.~L.~W., \& {Herbst}, E. 2008, \apj, 682, 283,
  \dodoi{10.1086/588035}

\bibitem[{{Graninger} {et~al.}(2016){Graninger}, {Wilkins}, \&
  {{\"O}berg}}]{Graninger16}
{Graninger}, D.~M., {Wilkins}, O.~H., \& {{\"O}berg}, K.~I. 2016, \apj, 819,
  140, \dodoi{10.3847/0004-637X/819/2/140}

\bibitem[{{Harada} \& {Herbst}(2008)}]{Harada08}
{Harada}, N., \& {Herbst}, E. 2008, \apj, 685, 272, \dodoi{10.1086/590468}

\bibitem[{{Hasegawa} \& {Herbst}(1993)}]{Hasegawa93}
{Hasegawa}, T.~I., \& {Herbst}, E. 1993, \mnras, 261, 83

\bibitem[{{Hassel} {et~al.}(2011){Hassel}, {Harada}, \& {Herbst}}]{Hassel11}
{Hassel}, G.~E., {Harada}, N., \& {Herbst}, E. 2011, \apj, 743, 182,
  \dodoi{10.1088/0004-637X/743/2/182}

\bibitem[{{Hassel} {et~al.}(2008){Hassel}, {Herbst}, \& {Garrod}}]{Hassel08}
{Hassel}, G.~E., {Herbst}, E., \& {Garrod}, R.~T. 2008, \apj, 681, 1385,
  \dodoi{10.1086/588185}

\bibitem[{{Herbst} \& {Leung}(1989)}]{Herbst89}
{Herbst}, E., \& {Leung}, C.~M. 1989, \apjs, 69, 271, \dodoi{10.1086/191314}

\bibitem[{{Higuchi} {et~al.}(2018){Higuchi}, {Sakai}, {Watanabe},
  {L{\'o}pez-Sepulcre}, {Yoshida}, {Oya}, {Imai}, {Zhang}, {Ceccarelli},
  {Lefloch}, {Codella}, {Bachiller}, {Hirota}, {Sakai}, \&
  {Yamamoto}}]{Higuchi18}
{Higuchi}, A.~E., {Sakai}, N., {Watanabe}, Y., {et~al.} 2018, \apjs, 236, 52,
  \dodoi{10.3847/1538-4365/aabfe9}

\bibitem[{{Hincelin} {et~al.}(2015){Hincelin}, {Chang}, \&
  {Herbst}}]{Hincelin15}
{Hincelin}, U., {Chang}, Q., \& {Herbst}, E. 2015, \aap, 574, A24,
  \dodoi{10.1051/0004-6361/201424807}

\bibitem[{{Hirota} {et~al.}(2009){Hirota}, {Ohishi}, \& {Yamamoto}}]{Hirota09}
{Hirota}, T., {Ohishi}, M., \& {Yamamoto}, S. 2009, \apj, 699, 585,
  \dodoi{10.1088/0004-637X/699/1/585}

\bibitem[{{Hirota} {et~al.}(2010){Hirota}, {Sakai}, \& {Yamamoto}}]{Hirota10}
{Hirota}, T., {Sakai}, N., \& {Yamamoto}, S. 2010, \apj, 720, 1370,
  \dodoi{10.1088/0004-637X/720/2/1370}

\bibitem[{{Hocuk} {et~al.}(2016){Hocuk}, {Cazaux}, {Spaans}, \&
  {Caselli}}]{Hocuk16}
{Hocuk}, S., {Cazaux}, S., {Spaans}, M., \& {Caselli}, P. 2016, MNRAS, 456,
  2586, \dodoi{10.1093/mnras/stv2817}

\bibitem[{{Hocuk} {et~al.}(2017){Hocuk}, {Sz{\H u}cs}, {Caselli}, {Cazaux},
  {Spaans}, \& {Esplugues}}]{Hocuk17}
{Hocuk}, S., {Sz{\H u}cs}, L., {Caselli}, P., {et~al.} 2017, \aap, 604, A58,
  \dodoi{10.1051/0004-6361/201629944}

\bibitem[{{Imai} {et~al.}(2016){Imai}, {Sakai}, {Oya}, {L{\'o}pez-Sepulcre},
  {Watanabe}, {Ceccarelli}, {Lefloch}, {Caux}, {Vastel}, {Kahane}, {Sakai},
  {Hirota}, {Aikawa}, \& {Yamamoto}}]{Imai16}
{Imai}, M., {Sakai}, N., {Oya}, Y., {et~al.} 2016, \apjl, 830, L37,
  \dodoi{10.3847/2041-8205/830/2/L37}

\bibitem[{{Ivlev} {et~al.}(2015){Ivlev}, {Padovani}, {Galli}, \&
  {Caselli}}]{Ivlev15p}
{Ivlev}, A.~V., {Padovani}, M., {Galli}, D., \& {Caselli}, P. 2015, \apj, 812,
  135, \dodoi{10.1088/0004-637X/812/2/135}

\bibitem[{{J{\o}rgensen} {et~al.}(2002){J{\o}rgensen}, {Sch{\"o}ier}, \& {van
  Dishoeck}}]{Jorgensen02}
{J{\o}rgensen}, J.~K., {Sch{\"o}ier}, F.~L., \& {van Dishoeck}, E.~F. 2002,
  \aap, 389, 908, \dodoi{10.1051/0004-6361:20020681}

\bibitem[{{J{\o}rgensen} {et~al.}(2013){J{\o}rgensen}, {Visser}, {Sakai},
  {Bergin}, {Brinch}, {Harsono}, {Lindberg}, {van Dishoeck}, {Yamamoto},
  {Bisschop}, \& {Persson}}]{Jorgensen13}
{J{\o}rgensen}, J.~K., {Visser}, R., {Sakai}, N., {et~al.} 2013, \apjl, 779,
  L22, \dodoi{10.1088/2041-8205/779/2/L22}

\bibitem[{{Kalv{\= a}ns}(2015{\natexlab{a}})}]{K15apj1}
{Kalv{\= a}ns}, J. 2015{\natexlab{a}}, \apj, 803, 52,
  \dodoi{10.1088/0004-637X/803/2/52}

\bibitem[{{Kalv{\= a}ns}(2015{\natexlab{b}})}]{K15apj2}
---. 2015{\natexlab{b}}, \apj, 806, 196, \dodoi{10.1088/0004-637X/806/2/196}

\bibitem[{{Kalv{\= a}ns}(2018{\natexlab{a}})}]{K18aps}
---. 2018{\natexlab{a}}, \apjs, 239, 6 (Paper~II),
  \dodoi{10.3847/1538-4365/aae527}

\bibitem[{{Kalv{\= a}ns}(2018{\natexlab{b}})}]{K18mn}
---. 2018{\natexlab{b}}, \mnras, 478, 2753, \dodoi{10.1093/mnras/sty1172}

\bibitem[{{Kalv{\= a}ns} \& {Kalnin}(2019)}]{KK19}
{Kalv{\= a}ns}, J., \& {Kalnin}, J.~R. 2019, \mnras, 486, 2050,
  \dodoi{10.1093/mnras/stz1010}

\bibitem[{{Kalv{\= a}ns} {et~al.}(2017){Kalv{\= a}ns}, {Shmeld}, {Kalnin}, \&
  {Hocuk}}]{K17}
{Kalv{\= a}ns}, J., {Shmeld}, I., {Kalnin}, J.~R., \& {Hocuk}, S. 2017, \mnras,
  467, 1763, \dodoi{10.1093/mnras/stx174}

\bibitem[{{Kalv{\={a}}ns} \& {Kalnin}(2020{\natexlab{a}})}]{KK20}
{Kalv{\={a}}ns}, J., \& {Kalnin}, J.~R. 2020{\natexlab{a}}, \aap, 633, A97,
  \dodoi{10.1051/0004-6361/201936471}

\bibitem[{{Kalv{\={a}}ns} \& {Kalnin}(2020{\natexlab{b}})}]{KK20ii}
---. 2020{\natexlab{b}}, \aap, 641, A49, \dodoi{10.1051/0004-6361/202037906}

\bibitem[{{Keto} \& {Caselli}(2010)}]{Keto10}
{Keto}, E., \& {Caselli}, P. 2010, \mnras, 402, 1625,
  \dodoi{10.1111/j.1365-2966.2009.16033.x}

\bibitem[{{Laas} {et~al.}(2011){Laas}, {Garrod}, {Herbst}, \& {Widicus
  Weaver}}]{Laas11}
{Laas}, J.~C., {Garrod}, R.~T., {Herbst}, E., \& {Widicus Weaver}, S.~L. 2011,
  \apj, 728, 71, \dodoi{10.1088/0004-637X/728/1/71}

\bibitem[{{Launhardt} {et~al.}(2013){Launhardt}, {Stutz}, {Schmiedeke},
  {Henning}, {Krause}, {Balog}, {Beuther}, {Birkmann}, {Hennemann},
  {Kainulainen}, {Khanzadyan}, {Linz}, {Lippok}, {Nielbock}, {Pitann}, {Ragan},
  {Risacher}, {Schmalzl}, {Shirley}, {Stecklum}, {Steinacker}, \&
  {Tackenberg}}]{Launhardt13}
{Launhardt}, R., {Stutz}, A.~M., {Schmiedeke}, A., {et~al.} 2013, \aap, 551,
  A98, \dodoi{10.1051/0004-6361/201220477}

\bibitem[{{Law} {et~al.}(2018){Law}, {{\"O}berg}, {Bergner}, \&
  {Graninger}}]{Law18}
{Law}, C.~J., {{\"O}berg}, K.~I., {Bergner}, J.~B., \& {Graninger}, D. 2018,
  \apj, 863, 88, \dodoi{10.3847/1538-4357/aacf9d}

\bibitem[{{Lee} {et~al.}(1996){Lee}, {Herbst}, {Pineau des Forets}, {Roueff},
  \& {Le Bourlot}}]{Lee96}
{Lee}, H.-H., {Herbst}, E., {Pineau des Forets}, G., {Roueff}, E., \& {Le
  Bourlot}, J. 1996, \aap, 311, 690

\bibitem[{{Lefloch} {et~al.}(2018){Lefloch}, {Bachiller}, {Ceccarelli},
  {Cernicharo}, {Codella}, {Fuente}, {Kahane}, {L{\'o}pez-Sepulcre}, {Tafalla},
  {Vastel}, {Caux}, {Gonz{\'a}lez-Garc{\'\i}a}, {Bianchi}, {G{\'o}mez-Ruiz},
  {Holdship}, {Mendoza}, {Ospina-Zamudio}, {Podio}, {Qu{\'e}nard}, {Roueff},
  {Sakai}, {Viti}, {Yamamoto}, {Yoshida}, {Favre}, {Monfredini},
  {Quiti{\'a}n-Lara}, {Marcelino}, {Boechat-Roberty}, \& {Cabrit}}]{Lefloch18}
{Lefloch}, B., {Bachiller}, R., {Ceccarelli}, C., {et~al.} 2018, \mnras, 477,
  4792, \dodoi{10.1093/mnras/sty937}

\bibitem[{{Leger} {et~al.}(1985){Leger}, {Jura}, \& {Omont}}]{Leger85}
{Leger}, A., {Jura}, M., \& {Omont}, A. 1985, \aap, 144, 147

\bibitem[{{Li} {et~al.}(2016){Li}, {Shen}, {Wang}, {Chen}, {Wu}, {Zhao},
  {Wang}, {Zuo}, {Fan}, {Hong}, {Jiang}, {Li}, {Liang}, {Ling}, {Liu}, {Qian},
  {Zhang}, {Zhong}, \& {Ye}}]{Li16}
{Li}, J., {Shen}, Z.-Q., {Wang}, J., {et~al.} 2016, \apj, 824, 136,
  \dodoi{10.3847/0004-637X/824/2/136}

\bibitem[{{Li} {et~al.}(2013){Li}, {Heays}, {Visser}, {Ubachs}, {Lewis},
  {Gibson}, \& {van Dishoeck}}]{Li13}
{Li}, X., {Heays}, A.~N., {Visser}, R., {et~al.} 2013, \aap, 555, A14,
  \dodoi{10.1051/0004-6361/201220625}

\bibitem[{{Lindberg} {et~al.}(2016){Lindberg}, {Charnley}, \&
  {Cordiner}}]{Lindberg16}
{Lindberg}, J.~E., {Charnley}, S.~B., \& {Cordiner}, M.~A. 2016, \apjl, 833,
  L14, \dodoi{10.3847/2041-8213/833/1/L14}

\bibitem[{{Lindberg} \& {J{\o}rgensen}(2012)}]{Lindberg12}
{Lindberg}, J.~E., \& {J{\o}rgensen}, J.~K. 2012, \aap, 548, A24,
  \dodoi{10.1051/0004-6361/201219603}

\bibitem[{{McElroy} {et~al.}(2013){McElroy}, {Walsh}, {Markwick}, {Cordiner},
  {Smith}, \& {Millar}}]{McElroy13}
{McElroy}, D., {Walsh}, C., {Markwick}, A.~J., {et~al.} 2013, \aap, 550, A36,
  \dodoi{10.1051/0004-6361/201220465}

\bibitem[{{Mookerjea} {et~al.}(2010){Mookerjea}, {Giesen}, {Stutzki},
  {Cernicharo}, {Goicoechea}, {de Luca}, {Bell}, {Gupta}, {Gerin}, {Persson},
  {Sonnentrucker}, {Makai}, {Black}, {Boulanger}, {Coutens}, {Dartois},
  {Encrenaz}, {Falgarone}, {Geballe}, {Godard}, {Goldsmith}, {Gry},
  {Hennebelle}, {Herbst}, {Hily-Blant}, {Joblin}, {Ka{\'z}mierczak},
  {Ko{\l}os}, {Kre{\l}owski}, {Lis}, {Martin-Pintado}, {Menten}, {Monje},
  {Pearson}, {Perault}, {Phillips}, {Plume}, {Salez}, {Schlemmer}, {Schmidt},
  {Teyssier}, {Vastel}, {Yu}, {Dieleman}, {G{\"u}sten}, {Honingh}, {Morris},
  {Roelfsema}, {Schieder}, {Tielens}, \& {Zmuidzinas}}]{Mookerjea10}
{Mookerjea}, B., {Giesen}, T., {Stutzki}, J., {et~al.} 2010, \aap, 521, L13,
  \dodoi{10.1051/0004-6361/201015095}

\bibitem[{{Mookerjea} {et~al.}(2012){Mookerjea}, {Hassel}, {Gerin}, {Giesen},
  {Stutzki}, {Herbst}, {Black}, {Goldsmith}, {Menten}, {Kre{\l}owski}, {De
  Luca}, {Csengeri}, {Joblin}, {Ka{\'z}mierczak}, {Schmidt}, {Goicoechea}, \&
  {Cernicharo}}]{Mookerjea12}
{Mookerjea}, B., {Hassel}, G.~E., {Gerin}, M., {et~al.} 2012, \aap, 546, A75,
  \dodoi{10.1051/0004-6361/201219287}

\bibitem[{{Nejad} {et~al.}(1990){Nejad}, {Williams}, \& {Charnley}}]{Nejad90}
{Nejad}, L.~A.~M., {Williams}, D.~A., \& {Charnley}, S.~B. 1990, \mnras, 246,
  183

\bibitem[{{{\"O}berg} {et~al.}(2008){{\"O}berg}, {Boogert}, {Pontoppidan},
  {Blake}, {Evans}, {Lahuis}, \& {van Dishoeck}}]{Oberg08}
{{\"O}berg}, K.~I., {Boogert}, A.~C.~A., {Pontoppidan}, K.~M., {et~al.} 2008,
  \apj, 678, 1032, \dodoi{10.1086/533432}

\bibitem[{{Ohashi} {et~al.}(1997){Ohashi}, {Hayashi}, {Ho}, \&
  {Momose}}]{Ohashi97}
{Ohashi}, N., {Hayashi}, M., {Ho}, P. T.~P., \& {Momose}, M. 1997, \apj, 475,
  211, \dodoi{10.1086/303533}

\bibitem[{{Oya}(2020)}]{Oya20}
{Oya}, Y. 2020, in IAU Symposium, Vol. 345, IAU Symposium, ed. B.~G.
  {Elmegreen}, L.~V. {T{\'o}th}, \& M.~{G{\"u}del}, 111--114,
  \dodoi{10.1017/S1743921318008372}

\bibitem[{{Oya} {et~al.}(2015){Oya}, {Sakai}, {Lefloch}, {L{\'o}pez-Sepulcre},
  {Watanabe}, {Ceccarelli}, \& {Yamamoto}}]{Oya15}
{Oya}, Y., {Sakai}, N., {Lefloch}, B., {et~al.} 2015, \apj, 812, 59,
  \dodoi{10.1088/0004-637X/812/1/59}

\bibitem[{{Oya} {et~al.}(2017){Oya}, {Sakai}, {Watanabe}, {Higuchi}, {Hirota},
  {L{\'o}pez-Sepulcre}, {Sakai}, {Aikawa}, {Ceccarelli}, {Lefloch}, {Caux},
  {Vastel}, {Kahane}, \& {Yamamoto}}]{Oya17}
{Oya}, Y., {Sakai}, N., {Watanabe}, Y., {et~al.} 2017, \apj, 837, 174,
  \dodoi{10.3847/1538-4357/aa6300}

\bibitem[{{Padovani} {et~al.}(2009){Padovani}, {Galli}, \&
  {Glassgold}}]{Padovani09}
{Padovani}, M., {Galli}, D., \& {Glassgold}, A.~E. 2009, \aap, 501, 619,
  \dodoi{10.1051/0004-6361/200911794}

\bibitem[{{Pan} {et~al.}(2017){Pan}, {Li}, {Chang}, {Qian}, {Bergin}, \&
  {Wang}}]{Pan17}
{Pan}, Z., {Li}, D., {Chang}, Q., {et~al.} 2017, \apj, 836, 194,
  \dodoi{10.3847/1538-4357/aa5c33}

\bibitem[{{Plummer}(1911)}]{Plummer11}
{Plummer}, H.~C. 1911, \mnras, 71, 460

\bibitem[{{Rawlings} {et~al.}(2002){Rawlings}, {Hartquist}, {Williams}, \&
  {Falle}}]{Rawlings02}
{Rawlings}, J.~M.~C., {Hartquist}, T.~W., {Williams}, D.~A., \& {Falle},
  S.~A.~E.~G. 2002, \aap, 391, 681, \dodoi{10.1051/0004-6361:20020825}

\bibitem[{{Sakai} {et~al.}(2009{\natexlab{a}}){Sakai}, {Sakai}, {Hirota},
  {Burton}, \& {Yamamoto}}]{Sakai09lupus}
{Sakai}, N., {Sakai}, T., {Hirota}, T., {Burton}, M., \& {Yamamoto}, S.
  2009{\natexlab{a}}, \apj, 697, 769, \dodoi{10.1088/0004-637X/697/1/769}

\bibitem[{{Sakai} {et~al.}(2008{\natexlab{a}}){Sakai}, {Sakai}, {Hirota}, \&
  {Yamamoto}}]{Sakai08}
{Sakai}, N., {Sakai}, T., {Hirota}, T., \& {Yamamoto}, S. 2008{\natexlab{a}},
  \apj, 672, 371, \dodoi{10.1086/523635}

\bibitem[{{Sakai} {et~al.}(2009{\natexlab{b}}){Sakai}, {Sakai}, {Hirota}, \&
  {Yamamoto}}]{Sakai09deut}
---. 2009{\natexlab{b}}, \apj, 702, 1025, \dodoi{10.1088/0004-637X/702/2/1025}

\bibitem[{{Sakai} {et~al.}(2010){Sakai}, {Sakai}, {Hirota}, \&
  {Yamamoto}}]{Sakai10}
---. 2010, \apj, 722, 1633, \dodoi{10.1088/0004-637X/722/2/1633}

\bibitem[{{Sakai} {et~al.}(2007){Sakai}, {Sakai}, {Osamura}, \&
  {Yamamoto}}]{Sakai07}
{Sakai}, N., {Sakai}, T., {Osamura}, Y., \& {Yamamoto}, S. 2007, \apjl, 667,
  L65, \dodoi{10.1086/521979}

\bibitem[{{Sakai} {et~al.}(2008{\natexlab{b}}){Sakai}, {Sakai}, \&
  {Yamamoto}}]{Sakai08-}
{Sakai}, N., {Sakai}, T., \& {Yamamoto}, S. 2008{\natexlab{b}}, \apjl, 673,
  L71, \dodoi{10.1086/527376}

\bibitem[{{Sakai} {et~al.}(2012){Sakai}, {Shirley}, {Sakai}, {Hirota},
  {Watanabe}, \& {Yamamoto}}]{Sakai12}
{Sakai}, N., {Shirley}, Y.~L., {Sakai}, T., {et~al.} 2012, \apjl, 758, L4,
  \dodoi{10.1088/2041-8205/758/1/L4}

\bibitem[{{Sakai} \& {Yamamoto}(2013)}]{Sakai13}
{Sakai}, N., \& {Yamamoto}, S. 2013, Chemical Reviews, 113, 8981,
  \dodoi{10.1021/cr4001308}

\bibitem[{{Sakai} {et~al.}(2014{\natexlab{a}}){Sakai}, {Oya}, {Sakai},
  {Watanabe}, {Hirota}, {Ceccarelli}, {Kahane}, {Lopez-Sepulcre}, {Lefloch},
  {Vastel}, {Bottinelli}, {Caux}, {Coutens}, {Aikawa}, {Takakuwa}, {Ohashi},
  {Yen}, \& {Yamamoto}}]{Sakai14}
{Sakai}, N., {Oya}, Y., {Sakai}, T., {et~al.} 2014{\natexlab{a}}, \apjl, 791,
  L38, \dodoi{10.1088/2041-8205/791/2/L38}

\bibitem[{{Sakai} {et~al.}(2014{\natexlab{b}}){Sakai}, {Sakai}, {Hirota},
  {Watanabe}, {Ceccarelli}, {Kahane}, {Bottinelli}, {Caux}, {Demyk}, {Vastel},
  {Coutens}, {Taquet}, {Ohashi}, {Takakuwa}, {Yen}, {Aikawa}, \&
  {Yamamoto}}]{Sakai14nat}
{Sakai}, N., {Sakai}, T., {Hirota}, T., {et~al.} 2014{\natexlab{b}}, \nat, 507,
  78, \dodoi{10.1038/nature13000}

\bibitem[{{Sakai} {et~al.}(2016){Sakai}, {Oya}, {L{\'o}pez-Sepulcre},
  {Watanabe}, {Sakai}, {Hirota}, {Aikawa}, {Ceccarelli}, {Lefloch}, {Caux},
  {Vastel}, {Kahane}, \& {Yamamoto}}]{Sakai16}
{Sakai}, N., {Oya}, Y., {L{\'o}pez-Sepulcre}, A., {et~al.} 2016, \apjl, 820,
  L34, \dodoi{10.3847/2041-8205/820/2/L34}

\bibitem[{{Sakai} {et~al.}(2017){Sakai}, {Oya}, {Higuchi}, {Aikawa}, {Hanawa},
  {Ceccarelli}, {Lefloch}, {L{\'o}pez-Sepulcre}, {Watanabe}, {Sakai}, {Hirota},
  {Caux}, {Vastel}, {Kahane}, \& {Yamamoto}}]{Sakai17}
{Sakai}, N., {Oya}, Y., {Higuchi}, A.~E., {et~al.} 2017, \mnras, 467, L76,
  \dodoi{10.1093/mnrasl/slx002}

\bibitem[{{Sandford} {et~al.}(1988){Sandford}, {Allamandola}, {Tielens}, \&
  {Valero}}]{Sandford88}
{Sandford}, S.~A., {Allamandola}, L.~J., {Tielens}, A.~G.~G.~M., \& {Valero},
  G.~J. 1988, \apj, 329, 498, \dodoi{10.1086/166395}

\bibitem[{{Saul} {et~al.}(2015){Saul}, {Tothill}, \& {Purcell}}]{Saul15}
{Saul}, M., {Tothill}, N.~F.~H., \& {Purcell}, C.~R. 2015, \apj, 798, 36,
  \dodoi{10.1088/0004-637X/798/1/36}

\bibitem[{{Semenov} {et~al.}(2010){Semenov}, {Hersant}, {Wakelam}, {Dutrey},
  {Chapillon}, {Guilloteau}, {Henning}, {Launhardt}, {Pi{\'e}tu}, \&
  {Schreyer}}]{Semenov10}
{Semenov}, D., {Hersant}, F., {Wakelam}, V., {et~al.} 2010, \aap, 522, A42,
  \dodoi{10.1051/0004-6361/201015149}

\bibitem[{{Sicilia-Aguilar} {et~al.}(2019){Sicilia-Aguilar}, {Patel}, {Fang},
  {Roccatagliata}, {Getman}, \& {Goldsmith}}]{Sicilia19}
{Sicilia-Aguilar}, A., {Patel}, N., {Fang}, M., {et~al.} 2019, \aap, 622, A118,
  \dodoi{10.1051/0004-6361/201833207}

\bibitem[{{Spezzano} {et~al.}(2016){Spezzano}, {Bizzocchi}, {Caselli}, {Harju},
  \& {Br{\"u}nken}}]{Spezzano16}
{Spezzano}, S., {Bizzocchi}, L., {Caselli}, P., {Harju}, J., \& {Br{\"u}nken},
  S. 2016, \aap, 592, L11, \dodoi{10.1051/0004-6361/201628652}

\bibitem[{{Strong} \& {Moskalenko}(1998)}]{Strong98}
{Strong}, A.~W., \& {Moskalenko}, I.~V. 1998, \apj, 509, 212,
  \dodoi{10.1086/306470}

\bibitem[{{Takakuwa} {et~al.}(2011){Takakuwa}, {Ohashi}, \&
  {Aikawa}}]{Takakuwa11}
{Takakuwa}, S., {Ohashi}, N., \& {Aikawa}, Y. 2011, \apj, 728, 101,
  \dodoi{10.1088/0004-637X/728/2/101}

\bibitem[{{Takano} {et~al.}(1990){Takano}, {Suzuki}, {Ohishi}, {Ishikawa},
  {Kaifu}, {Hirahara}, \& {Masuda}}]{Takano90}
{Takano}, S., {Suzuki}, H., {Ohishi}, M., {et~al.} 1990, \apjl, 361, L15,
  \dodoi{10.1086/185816}

\bibitem[{{Taniguchi} {et~al.}(2019){Taniguchi}, {Herbst}, {Caselli},
  {Paulive}, {Maffucci}, \& {Saito}}]{Taniguchi19}
{Taniguchi}, K., {Herbst}, E., {Caselli}, P., {et~al.} 2019, \apj, 881, 57,
  \dodoi{10.3847/1538-4357/ab2d9e}

\bibitem[{{Taniguchi} {et~al.}(2018{\natexlab{a}}){Taniguchi}, {Miyamoto},
  {Saito}, {Sanhueza}, {Shimoikura}, {Dobashi}, {Nakamura}, \&
  {Ozeki}}]{Taniguchi18poly}
{Taniguchi}, K., {Miyamoto}, Y., {Saito}, M., {et~al.} 2018{\natexlab{a}},
  \apj, 866, 32, \dodoi{10.3847/1538-4357/aadd0c}

\bibitem[{{Taniguchi} {et~al.}(2018{\natexlab{b}}){Taniguchi}, {Saito},
  {Majumdar}, {Shimoikura}, {Dobashi}, {Ozeki}, {Nakamura}, {Hirota},
  {Minamidani}, {Miyamoto}, \& {Kaneko}}]{Taniguchi18mas}
{Taniguchi}, K., {Saito}, M., {Majumdar}, L., {et~al.} 2018{\natexlab{b}},
  \apj, 866, 150, \dodoi{10.3847/1538-4357/aade97}

\bibitem[{{Taquet} {et~al.}(2014){Taquet}, {Charnley}, \&
  {Sipil{\"a}}}]{Taquet14}
{Taquet}, V., {Charnley}, S.~B., \& {Sipil{\"a}}, O. 2014, \apj, 791, 1,
  \dodoi{10.1088/0004-637X/791/1/1}

\bibitem[{{Thi} {et~al.}(2010){Thi}, {Woitke}, \& {Kamp}}]{Thi10}
{Thi}, W.-F., {Woitke}, P., \& {Kamp}, I. 2010, \mnras, 407, 232,
  \dodoi{10.1111/j.1365-2966.2009.16162.x}

\bibitem[{{Tokudome} {et~al.}(2013){Tokudome}, {Sakai}, {Sakai}, {Takano},
  {Yamamoto}, \& {NRO Line Survey Project Members}}]{Tokudome13}
{Tokudome}, T., {Sakai}, N., {Sakai}, T., {et~al.} 2013, in Astronomical
  Society of the Pacific Conference Series, Vol. 476, New Trends in Radio
  Astronomy in the ALMA Era: The 30th Anniversary of Nobeyama Radio
  Observatory, ed. R.~{Kawabe}, N.~{Kuno}, \& S.~{Yamamoto}, 355

\bibitem[{{Wakelam} \& {Herbst}(2008)}]{Wakelam08}
{Wakelam}, V., \& {Herbst}, E. 2008, \apj, 680, 371, \dodoi{10.1086/587734}

\bibitem[{{Wang} {et~al.}(2019){Wang}, {Chang}, \& {Wang}}]{Wang19}
{Wang}, Y., {Chang}, Q., \& {Wang}, H. 2019, \aap, 622, A185,
  \dodoi{10.1051/0004-6361/201834276}

\bibitem[{{Watanabe} {et~al.}(2012){Watanabe}, {Sakai}, {Lindberg},
  {J{\o}rgensen}, {Bisschop}, \& {Yamamoto}}]{Watanabe12}
{Watanabe}, Y., {Sakai}, N., {Lindberg}, J.~E., {et~al.} 2012, \apj, 745, 126,
  \dodoi{10.1088/0004-637X/745/2/126}

\bibitem[{{Whittet} {et~al.}(2001){Whittet}, {Gerakines}, {Hough}, \&
  {Shenoy}}]{Whittet01}
{Whittet}, D.~C.~B., {Gerakines}, P.~A., {Hough}, J.~H., \& {Shenoy}, S.~S.
  2001, \apj, 547, 872, \dodoi{10.1086/318421}

\bibitem[{{Whittet} {et~al.}(2007){Whittet}, {Shenoy}, {Bergin}, {Chiar},
  {Gerakines}, {Gibb}, {Melnick}, \& {Neufeld}}]{Whittet07}
{Whittet}, D.~C.~B., {Shenoy}, S.~S., {Bergin}, E.~A., {et~al.} 2007, \apj,
  655, 332, \dodoi{10.1086/509772}

\bibitem[{{Whitworth} \& {Ward-Thompson}(2001)}]{Whitworth01}
{Whitworth}, A.~P., \& {Ward-Thompson}, D. 2001, \apj, 547, 317,
  \dodoi{10.1086/318373}

\bibitem[{{Wu} {et~al.}(2019{\natexlab{a}}){Wu}, {Lin}, {Liu}, {Chen}, {Liu},
  {Zhang}, {Ju}, {Yuan}, {Wang}, {Shen}, {Kim}, {Qin}, {Li}, {Liu}, {Zhang},
  {Xu}, \& {Liu}}]{Wu19aa}
{Wu}, Y., {Lin}, L., {Liu}, X., {et~al.} 2019{\natexlab{a}}, \aap, 627, A162,
  \dodoi{10.1051/0004-6361/201834184}

\bibitem[{{Wu} {et~al.}(2019{\natexlab{b}}){Wu}, {Liu}, {Chen}, {Lin}, {Yuan},
  {Zhang}, {Liu}, {Shen}, {Li}, {Wang}, {Qin}, {Kim}, {Liu}, {Zhu}, {Madones},
  {Inostroza}, {Henkel}, {Zhang}, {Li}, {Esimbek}, \& {Liu}}]{Wu19mn}
{Wu}, Y., {Liu}, X., {Chen}, X., {et~al.} 2019{\natexlab{b}}, \mnras, 488, 495,
  \dodoi{10.1093/mnras/stz1498}

\bibitem[{{Yoshida} {et~al.}(2019){Yoshida}, {Sakai}, {Nishimura}, {Tokudome},
  {Watanabe}, {Sakai}, {Takano}, \& {Yamamoto}}]{Yoshida19}
{Yoshida}, K., {Sakai}, N., {Nishimura}, Y., {et~al.} 2019, \pasj, 71, S18,
  \dodoi{10.1093/pasj/psy136}

\end{thebibliography}
\bibliographystyle{aasjournal}

\end{document}